\documentclass{aa}  
\usepackage{txfonts}
\usepackage{xcolor}
\usepackage{gensymb}
\usepackage{natbib}
\usepackage{longtable}
\usepackage[normalem]{ulem}

\def \1s{$1\,\sigma$}

\def \t0{T$_0$}

\def \spirou{SPIRou}
\def \hd189{HD~189733}
\begin{document} 
\title{Early science with SPIRou: near-infrared radial velocity and spectropolarimetry of the planet-hosting star \hd189\ \thanks{ Based on observations obtained at the Canada-France-Hawaii Telescope (CFHT) which is operated from the summit of Maunakea by the National Research Council of Canada, the Institut National des Sciences de l'Univers of the Centre National de la Recherche Scientifique of France, and the University of Hawaii. Based on observations obtained with SPIRou, an international project led by Institut de Recherche en Astrophysique et Plan\'etologie, Toulouse, France.}}
   \author{Moutou C., 
          \inst{1}
          \and
Dalal, S.\inst{2} \and
Donati, J.-F.\inst{1} \and
Martioli, E.\inst{2,9} \and
Folsom, C.P.\inst{1} \and
 Artigau, \'E.\inst{3} \and
 Boisse, I.\inst{4} \and
 Bouchy, F.\inst{5} \and
 Carmona, A.\inst{6}
 Cook, N.J.\inst{3} \and
 Delfosse, X.\inst{6} \and
 Doyon, R.\inst{3} \and
 Fouqu\'e, P.\inst{7,1} \and
 Gaisn\'e, G.\inst{5} \and
 H\'ebrard, G.\inst{2,10} \and
 Hobson, M.\inst{4} \and
 Klein, B.\inst{1} \and
 Lecavelier des Etangs, A.\inst{2} \and
 Morin, J.\inst{8}
          }
\institute{
\inst{1} Univ. de Toulouse, CNRS, IRAP, 14 avenue Belin, 31400 Toulouse, France, \email{claire.moutou@irap.omp.eu} \\
\inst{2}  Institut d'Astrophysique de Paris, UMR7095 CNRS, Universit\'e Pierre \& Marie Curie, 98bis boulevard Arago, 75014 Paris, France \\
\inst{3} Universit\'e de Montr\'eal, D\'epartement de Physique, IREX, Montr\'eal, QC, H3C 3J7, Canada\\
\inst{4} Aix-Marseille   Université,   CNRS,   CNES,   LAM   (Laboratoire d’Astrophysique de Marseille),  Marseille, France \\
\inst{5} Observatoire de Gen\`eve, Universit\'e de Gen\`eve, 1290 Sauverny, Switzerland\\
\inst{6} Univ. Grenoble Alpes, CNRS, IPAG, 38000 Grenoble, France\\
\inst{7} Canada-France-Hawaii Telescope, CNRS, 96743 Kamuela, Hawaii, USA\\
\inst{8}  Universit\'e de Montpellier, CNRS, LUPM,34095 Montpellier, France\\
\inst{9} Laborat\'{o}rio Nacional de Astrof\'{i}sica, Rua Estados Unidos 154, 37504-364, Itajub\'{a} - MG, Brazil\\
\inst{10} Observatoire de Haute Provence, St Michel l'Observatoire, France}
\date{Received April 7, 2020; accepted June 29, 2020}

\abstract
{\spirou\ is the newest spectropolarimeter and high-precision velocimeter that has recently been installed at the Canada-France-Hawaii Telescope on Maunakea, Hawaii. It operates in the near-infrared and simultaneously covers the 0.98-2.35\,$\mu$m domain at high spectral resolution. \spirou\ is optimized for exoplanet search and characterization with the radial-velocity technique, and for polarization measurements in stellar lines and subsequent magnetic field studies. The host of the transiting hot Jupiter \hd189\,b has been observed during early science runs.  We present the first near-infrared spectropolarimetric observations of the planet-hosting star as well as the stellar radial velocities as measured by \spirou\ throughout the planetary orbit and two transit sequences.  The planetary orbit and Rossiter-McLaughlin anomaly are both investigated and modeled.
The orbital parameters and obliquity are all compatible with the values found in the optical.
The obtained radial-velocity precision is compatible with about twice the photon-noise estimates for a K2 star under these conditions. The additional scatter around the orbit, of about 8 m/s, agrees with previous results that showed that the activity-induced scatter is the dominant factor.
We analyzed the polarimetric signal, Zeeman broadening, and chromospheric activity tracers such as the 1083nm HeI and the 1282nm Pa$\beta$ lines to  investigate stellar activity. 
First estimates of the average unsigned magnetic flux from the Zeeman broadening of the FeI lines give a magnetic flux of 290$\pm$58 G, and the large-scale longitudinal field shows typical values of a few Gauss. These observations illustrate the potential of \spirou\ for exoplanet characterization and magnetic and stellar activity studies.
}

\titlerunning{\hd189\ seen by \spirou}
\authorrunning{C. Moutou et al}
\maketitle

\section{Introduction}
The study and characterization of exoplanet systems has greatly benefited from the instrumental developments in high-resolution optical spectroscopy since the pioneer work by \citet{walker}, \citet{mayor}, and \citet{marcy} and in particular, from the improvement of simultaneous wavelength calibrations \citep{baranne} in the advent of precise radial-velocity (RV) measurements. While recent years have seen very interesting developments in the optical with instruments suh as ESPRESSO \citep{pepe2014} and EXPRES \citep{petersburg2020}, the strongest emphasis on precision-RV instrumentation has been in the near-infrared (NIR) domain, with CARMENES-NIR \citep{Beceril2017}, GIARPS \citep{Claudi2017}, IRD \citep{kotani2018}, HPF \citep{mahadevan2014}, CRIRES$+$ \citep{Dorn2016}, and \spirou\ \citep{donati2020}. This new branch of RV NIR spectrographs has started a new era, in which the characterization of exoplanet systems widens toward a spectral domain full of riches: not only can different planet hosts be reached (young stars, and the coolest red dwarfs), but different atomic and molecular species are accessible to extend their characterization. In the extrasolar systems that can be observed both at optical and NIR wavelengths, the relative amplitude of stellar activity and planet signatures can, for instance, be used to distinguish between the two phenomena \citep{hebrard2014,reiners2010,zechmeister2018}. Moreover, Zeeman broadening increases with wavelength, which offers new ways to explore the surface activity of host stars that plagues the radial velocity technique \citep{reiners2013}. In this era of new NIR spectrographs with precise RV measurements, \spirou\ offers the advantage of including the full $YJHK$ coverage (0.98-2.35\,$\mu$m) and a polarimeter to monitor the evolution of circular polarization in lines. This is an additional tool for stellar characterization \citep{hebrard2016}.

We study early \spirou\ science data of the well-known hot-Jupiter system \hd189. Although \spirou\ is optimized for stars cooler than K types, whiose wealth of information contained in the spectral domain is greater, it seemed interesting to revisit this system with a new instrument and study the spectroscopic content and behavior in the NIR domain. \hd189\ is a moderately active K2 dwarf star that is transited by a hot-Jupiter planet in a 2.218-day circular orbit \citep{bouchy2005}. It has a distant M-dwarf companion \citep{bakos2006}. Since the planet was discovered and the first spectroscopic transit observations were reported, it has been extensively studied with high-resolution optical spectrographs \citep[e.g.,][]{winn2006,bakos2006b,redfield2008, triaud09}. 

The stellar activity of the planet host \hd189\ has been observed with space-based photometry with MOST \citep{croll2007}, unveiling a 3\% flux modulation due to the rotation of contrasted features in the photosphere. Radial velocities obtained with SOPHIE in turn exhibit a stellar activity jitter amplitude of about 10\,m/s $rms$ when the planetary orbit is removed \citep{boisse2009}. 

In spectropolarimetry, the star was also extensively followed up with CFHT/ESPaDOnS and TBL/NARVAL for its strong magnetic activity. Features induced by star-planet interactions were likewise searched for \citep{moutou07,fares10,fares17}. The magnetic field of \hd189\ is characterized by a complex topology. The mean value of the total field is about 40\,G, and there is a strong contribution of the toroidal component \citep{fares17}. 

Recently, \hd189\ transits have been observed using CRIRES \citep{brogi2016,brogi2019}, GIANO \citep{brogi2018} and the NIR arm of CARMENES. These observations and previous ones in transmission spectroscopy led to the discovery of water, carbon monoxide, and helium in the extended atmosphere of the planet \citep{birkby2013,salz2018,alonsofloriano2019}. 

In this paper, we present data on the orbital signal, on the Rossiter-McLaughling radial-velocity anomaly, on the circularly polarized Zeeman signatures in photospheric atomic lines, on the average small-scale surface magnetic flux, and on Pa$\beta$ and HeI chromospheric lines of \hd189 as seen by \spirou. The transmission spectroscopy analysis will be presented in a forthcoming paper. In section \ref{data_acquisition_and_processing} we provide a short description of the instrument, the observing strategy, the data collection, and the methods used in the data analysis. In section~\ref{radial_velocity_analysis} we present our radial-velocity results. In section~\ref{stellar_features} we present our analyses of the stellar activity, of polarization features, and Zeeman broadening before we conclude in the last section.

\section{Data acquisition and processing \label{data_acquisition_and_processing}}
\subsection{\spirou}
\spirou\ is the new NIR spectropolarimeter and high-precision velocimeter installed in 2018 at the Cassegrain focus of the 3.6 m Canada-France-Hawaii Telescope atop Maunakea \citep{donati2020}. Its wide spectral range (0.98-2.35\,$\mu$m), high resolving power (about 70\,000), its temperature-controlled environment \citep{challita18}, and precise light injection devices \citep{pares12} are key for obtaining time series of stellar spectra with extreme stability \citep{donati2018}. The CFH telescope is guided with a system that includes a tip-tilt plate operating at a frequency of 50 Hz for the observations of \hd189 and a NIR imaging camera sensitive to $YJH$ bands \citep{barrick2018}. This system then allows the operator to position the star on the 1.29$\arcsec$ circular aperture. The achromatic polarimeter located in the Cassegrain unit is composed of two ZnSe quarter-wave rhombs and a Wollaston prism, splitting the beam into the two orthogonal states of the selected polarization (circular polarization was used in this study). The spectrograph is fed with three fluoride fibers: two science fibers collecting the light out of the polarimeter, and one calibration fiber for the spectral reference lamp \citep{micheau2018}. A short octogonal fiber in the spectrograph contributes to a better scrambling to increase the RV precision.  The \spirou\ cryogenic spectrograph is operated in vacuum, at 73 K, and is thermally regulated at a submillikelvin level \citep{reshetov2012}. The design of the echelle spectrograph is described in \citet{thibaut2012} and contains an R2 diffraction grating and a cross-disperser train consisting of three prisms. Finally, the detector used in \spirou\ is a 15-micron science-grade H4RG from Teledyne Systems with up-the-ramp readout mode; the  detector characteristics are presented in \citet{artigau2018}. The orders are tilted relative to the detector rows, providing a good numerical sampling of the resolution element. 

\spirou\ can be used in spectropolarimetric and spectroscopic mode. In both modes, the star light passes through the polarimeter that is located in the front end. In the spectroscopic mode, each exposure is acquired in a single rhomb position, while in the polarimetric mode, four sub-exposures with different rhomb configurations are necessary to retrieve each Stokes parameter, as usual in optical spectropolarimeters \citep{donati97}. Radial velocities of the star can be measured in both modes, 
and the same set of calibrations is used for the two \spirou\ modes.

\spirou\ is operated with an extensive calibration plan, including daytime and nighttime observations. Each afternoon a series of internal calibrations is run, including detector dark frames and instrumental background (fixed time of 600s), order localization and flat-field frames using a tungsten lamp, accurate wavelength solution using exposures from a stabilized Fabry-Perot \citep{perruchot} and uranium-neon hollow-cathod lamps. The Fabry-Perot is also used simultaneously to stellar observations in the reference channel, allowing a monitoring of the residual spectrograph shifts over a timescale of the night. Extensive tests have shown that the calibrated instrument internally delivers radial velocities at a precision level better than 0.5 m/s $rms$ over days. Nighttime observations for \spirou\ calibration plan include a sky background of 10~min, at least one nightly radial-velocity standard and a few bright hot stars to monitor the telluric absorption spectrum in various conditions (for a total 10~min of observing time, nightly). \spirou\ has demonstrated its sensitivity, resolving power, and radial-velocity precision on cool nonactive stars and was accepted as a guest instrument at CFHT in January 2019. The on-sky performances of \spirou\ are outlined in a forthcoming paper \citep{donati2020}. \spirou\ has been open to the CFHT community since February 2019.

\subsection{Observations}
\hd189\ was observed with \spirou\ between \hbox{July 2018} and \hbox{June 2019} as part of the science verification program and of the SPIRou Legacy Survey (SLS)\footnote{CFHT programs 18BD50 and 19AP40}. Fourty-eight and eighty-six spectra were obtained in polarimetric and in spectroscopic mode, respectively.  The polarimetric mode has been used in circular polarisation at a temporal sampling of about one Stokes V sequence per night when \spirou\ was operated at CFHT between July 28 and October 25, 2018, for a total of 12 visits. In addition to recovering the planetary orbit signature, this sparse data set allows us to probe the magnetic field of the stellar host, which is already known from optical measurements with ESPaDOnS and NARVAL at much denser time sampling \citep{moutou07,fares10,fares17}. It also gives the option of investigating how stellar activity in the NIR is modulated by the 12d stellar rotation period. The spectroscopic mode was used for the observation of two planetary transit sequences, on 21-22 September, 2018, and on 14-15 June, 2019. The first partial transit sequence was obtained during a commissioning night and the baseline before ingress is missing. The second transit sequence has a longer baseline with equal baselines before ingress and after egress as the in-transit duration. 

Table \ref{table:1} lists the nights with observations of \hd189 considered in this paper, and the average properties of the N spectra taken in each of these nights. For polarimetric measurements, 4 consecutive exposures were taken, while for each of the two transits 36 and 50 exposures were taken in spectroscopic mode. The signal-to-noise ratio (S/N) per 2.28-km/s pixel bin in the middle of the H band ranges from 110 to 280 for 245\,s exposures, with a median value of 230. Conditions were photometric during both transit sequences, with an average seeing estimated on the guiding images of 0.77$\arcsec$ and 0.80$\arcsec$. 

   \begin{table*}
      \caption[]{\spirou\ observations of \hd189. The exposure time corresponds to one exposure. The S/N is per pixel and per exposure at 1.7\,$\mu$m. The barycentric Julian date is shifted by -2450000. \spirou\ has been used in both polarimetric and spectroscopic modes ('Pol' and 'Sp'). Zero-point (ZP) indicates the mean CFHT/SKYPROBE extinction during the visit, when available, in magnitude \citep{cuillandre}. The last two columns "RV phn" and "RV $rms$" show the average photon noise of the sequence and the scatter observed at a given epoch over the N data points when the planet signal is removed, respectively.}
         \label{table:1}
         \begin{center}
         \begin{tabular}{llllllllllll}
            \hline
            \noalign{\smallskip}
            UT date   &  Mode & N & BJD & $\phi_{\rm 
            orb}$ &Texp & See & S/N & ZP& RV &RV phn& RV $rms$\\
                      &      &   & &   & [s] & [$\arcsec$]  & &&km/s&m/s&m/s\\
            \noalign{\smallskip}
            \hline
            \noalign{\smallskip}
2018-07-29 & Pol& 4& 8329.03 & 0.31&245.164 & 0.66 & 240& N/A &-2.4642&1.49&5.2\\
2018-07-30 & Pol& 4& 8329.98 & 0.74&245.164 & 0.74 & 254& 0.13&-2.0406&1.61&11.7\\
2018-08-01 & Pol& 4& 8331.95 & 0.63&245.164  & 0.75 & 214& 0.05&-2.0942&1.78&8.7\\
2018-08-02 & Pol& 4& 8333.02 & 0.11&245.164 & 1.05 & 205& 0.04&-2.4039& 1.85&8.1\\
2018-08-03 & Pol& 8& 8333.98 & 0.55&245.164 & 0.85 & 110& 1.39&-2.2192& 3.05&9.0\\
2018-08-04 & Pol& 4& 8335.04 & 0.46&245.164 & 0.90 & 170& 0.48&-2.3069& 2.08&6.9\\
2018-08-05 & Pol& 4& 8336.03 & 0.88&245.164 & 0.78 & 213& 0.03&-2.3144& 1.78&7.9\\
2018-08-06 & Pol& 4& 8336.96 & 0.46&245.164 & 0.83 & 215& 0.03&-2.1238& 1.77&5.1\\
2018-09-22 & Sp &36& 8383.84 & 0.00&250.736 & 0.77 & 250& 0.07&-2.3016&1.65&4.0\\
2018-09-23 & Pol& 4& 8384.83 & 0.46&245.164 & 0.52 & 280& 0.00&-2.2975& 1.52&3.4\\
2018-09-25 & Pol& 4& 8386.80 & 0.35&245.164 & 1.15 & 230& 0.00&-2.4244& 1.75&4.9\\
2018-09-26 & Pol& 4& 8387.87 & 0.83&245.164 & 0.47 & 260& 0.22&-2.0975&1.48&9.9\\
2018-10-26 & Pol& 4& 8417.75 & 0.30&245.164 & 1.25 & 249& 0.00&-2.4572&1.69&4.1\\
2019-06-15 & Sp &50& 8650.04 & 0.00&250.736 & 0.80 & 210& 0.02&-2.2605&1.76&6.0\\
             \noalign{\smallskip}
            \hline
         \end{tabular}
         \end{center}
   \end{table*}
    \begin{figure*}
   \centering
   \includegraphics[width=0.9\hsize]{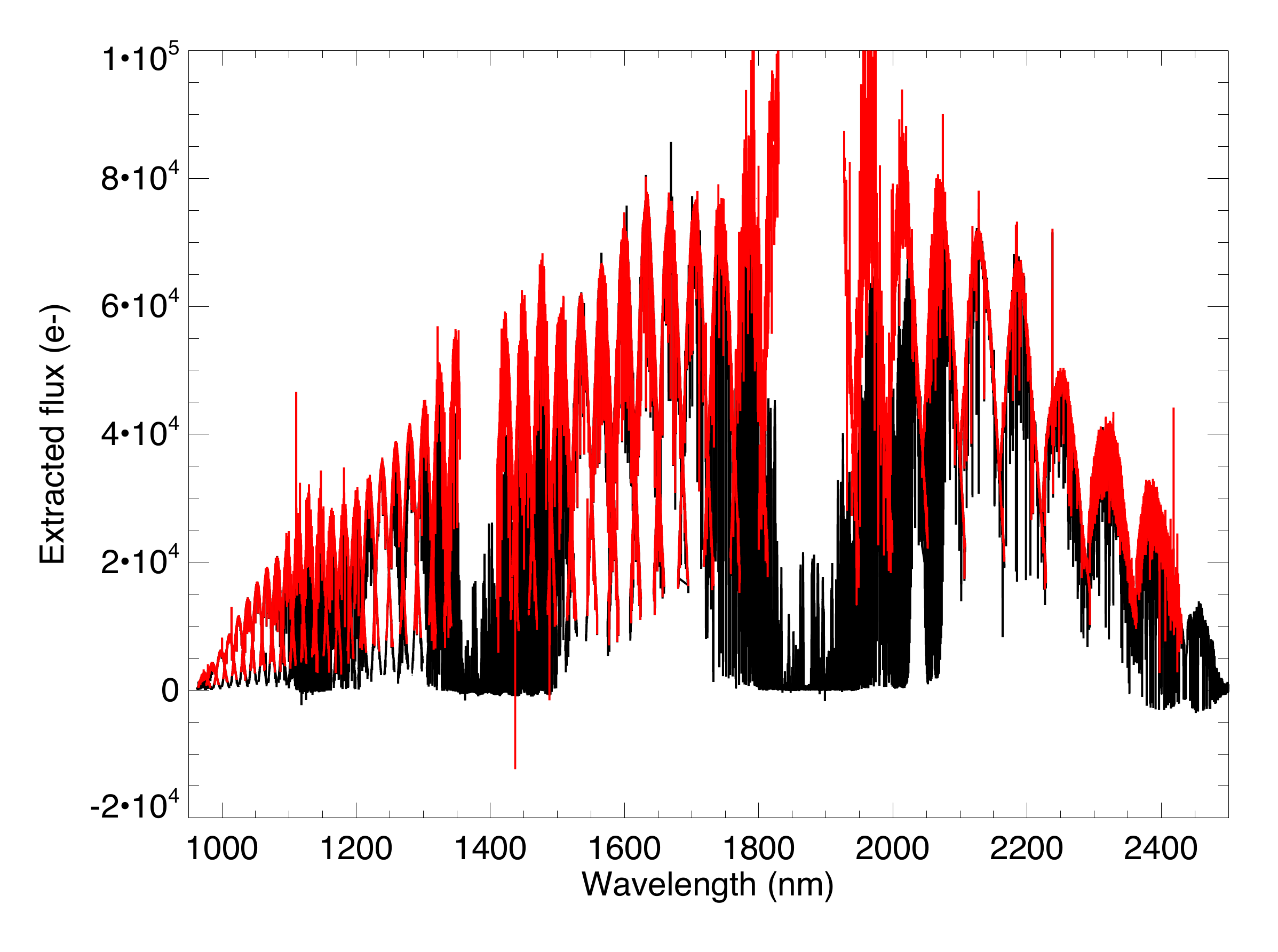}
   \includegraphics[width=0.9\hsize]{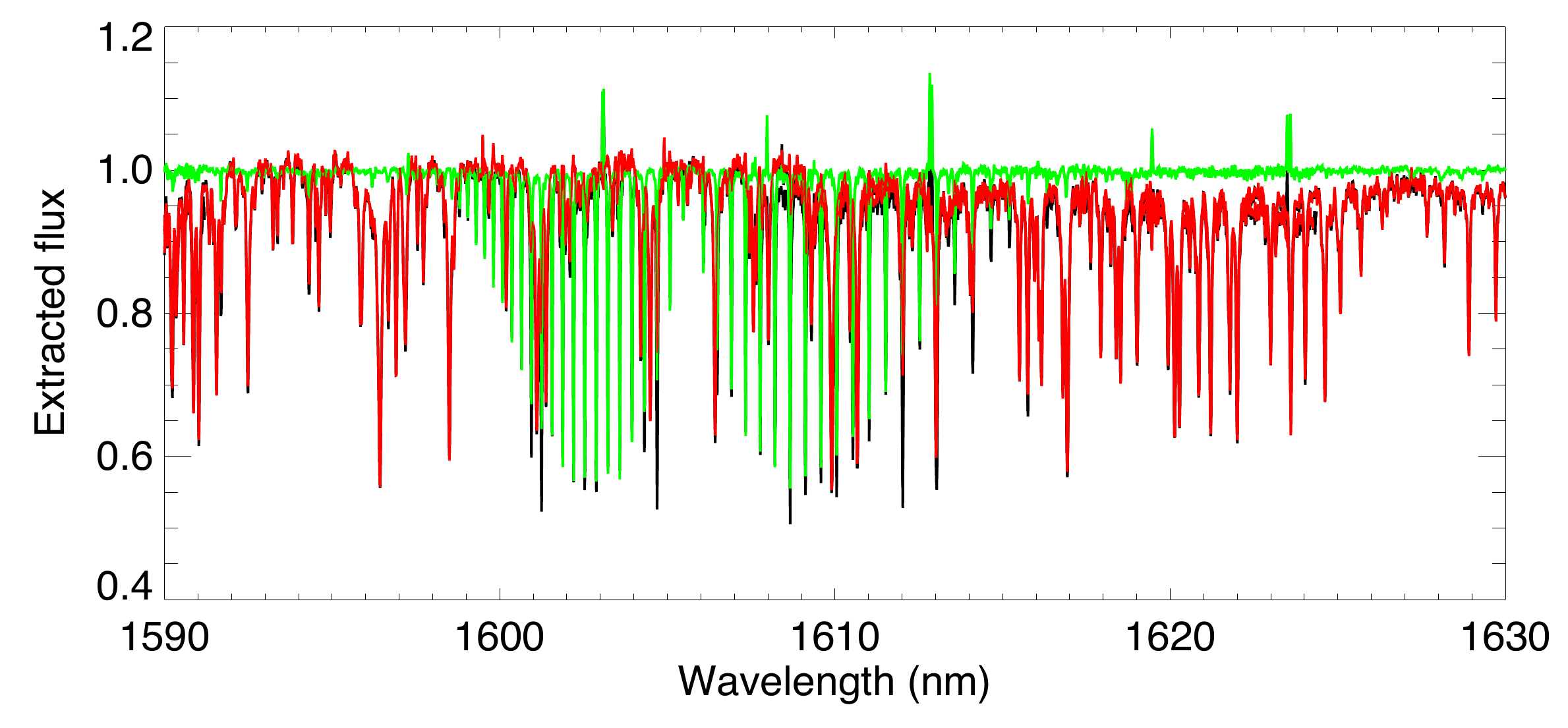}
   \caption{(Top) Example of the whole \spirou\ spectrum after pipeline extraction (black line) and after telluric correction (red line). (Bottom) Zoom into the $H$ band, showing the atmospheric spectrum derived from the PCA-based telluric correction. The removed telluric spectrum is shown in green; it contains absorption lines and most OH emission lines. The underlying black curve is the total extracted spectrum.              }
         \label{FigTellu}
   \end{figure*}

\subsection{Data processing}
The spectra were reduced with the \spirou\ data reduction system, APERO, 
using version 0.5.0. The nonlinearity of the detector was accounted for at the readout level, and the slope of individual pixels was calculated along the readout ramp within the linearity regime. The pipeline uses the optimal extraction method \citep{horne86} and the daily-updated calibration database to produce 2D spectra on which the radial-velocity measurement and (whenever relevant) the polarimetric calculation are performed. For RV measurements, the two science channels are extracted and treated together; for polarization, they are extracted individually and combined, as we describe in section 4.3. The Blaze function is estimated from the flat-field in each channel and removed from extracted spectra. The wavelength calibration is obtained from the combination of exposures using the UNe hollow-cathod lamp and the thermally controlled Fabry-P\'erot etalon (FP). The measurement of the FP cavity length is anchored on the absolute line positions of the UNe lamp, as originally proposed by \citet{bauer2015} and described in detail for \spirou\ by \citet{hobson2020}. The offset of the wavelength solution is then estimated for each exposure from the calibration of the simultaneous FP observation and the accurate calculation of the Barycentric Earth Radial Velocity (BERV).

Prior to cross-correlating the extracted spectra with a numerical mask to obtain the radial velocity, it is critical to correct for the telluric contamination that plagues the NIR domain. In the \spirou\ pipeline, we use the PCA-based correction technique that was first developed and tested on HARPS data, as described in \citet{artigau14}, which was then implemented for \spirou. Nightly observed telluric standards are collected into a library covering a widely sampled parameter space in atmospheric conditions, and this library is used to reconstruct with PCA the best-match telluric contamination spectrum of a given science observation. We used five principal components for telluric correction. 
Figure \ref{FigTellu} shows one example spectrum of \hd189\ that was extracted with the \spirou\ pipeline. In the top plot, the black dots show the full extracted spectrum, and the red spectrum was obtained after correction for telluric contamination, excluding the most opaque atmospheric bands. The blaze function was not removed at this stage. In the bottom plot, we show a small part of the $H$ band, after the blaze function was removed. In black and red we show the extracted stellar spectrum before and after correction of the most appropriate PCA-derived telluric spectrum, which is shown in green.

After the extracted spectra were corrected from the telluric contribution, the nominal RV calculation offered by the APERO pipeline was the cross-correlation function with a numerical mask, that was then fit by a Gaussian. The mean velocity of the Gaussian is the radial velocity. This is well adapted for slow rotators of K type such as \hd189 \citep{pepe2002,bouchy2005}, but has been shown to prevent the recovery of RV information content for fast rotators \citep[e.g.,][]{galland2005,yu2019} or M dwarfs in the optical  \citep[e.g.,][]{astudillo2015}.  

\subsection{Masks for cross-correlation}
The telluric-corrected spectra can be used to create a high-S/N template spectrum free of telluric lines, which is the result of combining all the spectra, excluding areas that are affected by a telluric line at the time of observation. We used the template of \hd189\ to produce a weighted mask that contains the central wavelength of the lines and a weight corresponding to their depth. The lines were first identified as a local minimum, more than 4$\sigma$ lower than the variation in the local continuum computed at a distance greater than the typical stellar line full-width at half-maximum (FWHM). This criterion rejected blended lines of similar depth. A Gaussian fit was then made around these local minima to determine the central position of the line and to reject features whose appearance was too different from that of a stellar line. See \citet{pepe2002} for a full description of the method of correlation with a weighted binary mask.

One difference between optical and NIR data is the role of residual telluric lines in the construction of the line list. Telluric lines show relative and variable motion with respect to the stellar lines, therefore the mask needs to be free of regions where the residual telluric correction still contaminates the spectrum and adds spurious RV variations.
We found that telluric lines deeper than 40\% leave significant residuals. Stellar lines that overlap telluric features with a depth larger than 40\% were then masked out for a BERV corresponding to the maximum range associated with \hd189. For shallower telluric lines, only the excess photon noise due to telluric residuals was included in the line weight (ranging from 1.5 times the photon noise for a line of 10\% depth to 7 times the photon noise for a line of 30\% depth). This mask-construction method is similar to method 3 described in \citet{figueira2016}. The produced K2 mask finally contained 1924 atomic and molecular lines in the whole \spirou\ domain. As an experiment, other partial masks were made from the original ones: a first set of three masks included only lines in the $J$, $H$, and $K$ bands from the K2 mask.

We then used the out-of-transit observations of the nights of September 21-22, 2018, and June 14-15, 2019, which were obtained in spectroscopic mode, to produce an alternative stellar mask. The line-list filtering makes use of a template spectrum obtained from the mean of all out-of-transit spectra, where we fit every line in the mask using a multiple Gaussian model to account for blends. After performing a standard least-squares fit, we compared the fitted central wavelength of each line with the mask values and  removed the lines in which the difference was greater than half of the FWHM, using FWHM$\sim$12\,km/s as the reference value. We also removed lines whose fitted amplitudes were lower than twice the mean flux error within the spectral region around the line. Finally, we removed lines in which the fitted FWHM lay outside an acceptable range of widths (a conservative range of 0.12 to 24 km/s was used here). 
The final K2-filtered mask contains 1405 lines.

A last set of masks was created from ATLAS9 stellar models \citep{castelli2003} and the VALD database, specifically, the model corresponding to an effective temperature of 5000\,K and log$g$ of 5.0. Originally, this mask contained 6102 atomic lines in the \spirou\ domain; when most opaque atmospheric bands at [1335-1490] and [1790-1994] were removed, 4121 lines remained. These constitute the Castelli5000 mask. From this line list, we then selected 1289 lines with the highest Land\'e factors (high Land\'e with a factor larger than 1.4 and a median factor of 1.57) and 1323 lines with the lowest Land\'e factors (low Land\'e with a factor smaller than 1.1 and a median factor of 0.91). 
The Land\'e factor of each line affects the line broadening because of the unsigned magnetic field in active regions of the stellar surface. By selecting the lines with the highest Land\'e factor, one amplifies the effect of the magnetic field on the line broadening, and potentially, on the radial-velocity jitter. One could thus expect the cross-correlation function (CCF) of the high Land\'e mask to have a larger width and smaller depth than the CCF obtained with the low Land\'e mask in the presence of a strong magnetic field. The largest number of lines, those with average Land\'e factors of about the median 1.25, remains in this selection to enhance any contrasted behavior.\\

\begin{figure}
   \centering
   \includegraphics[width=0.9\hsize]{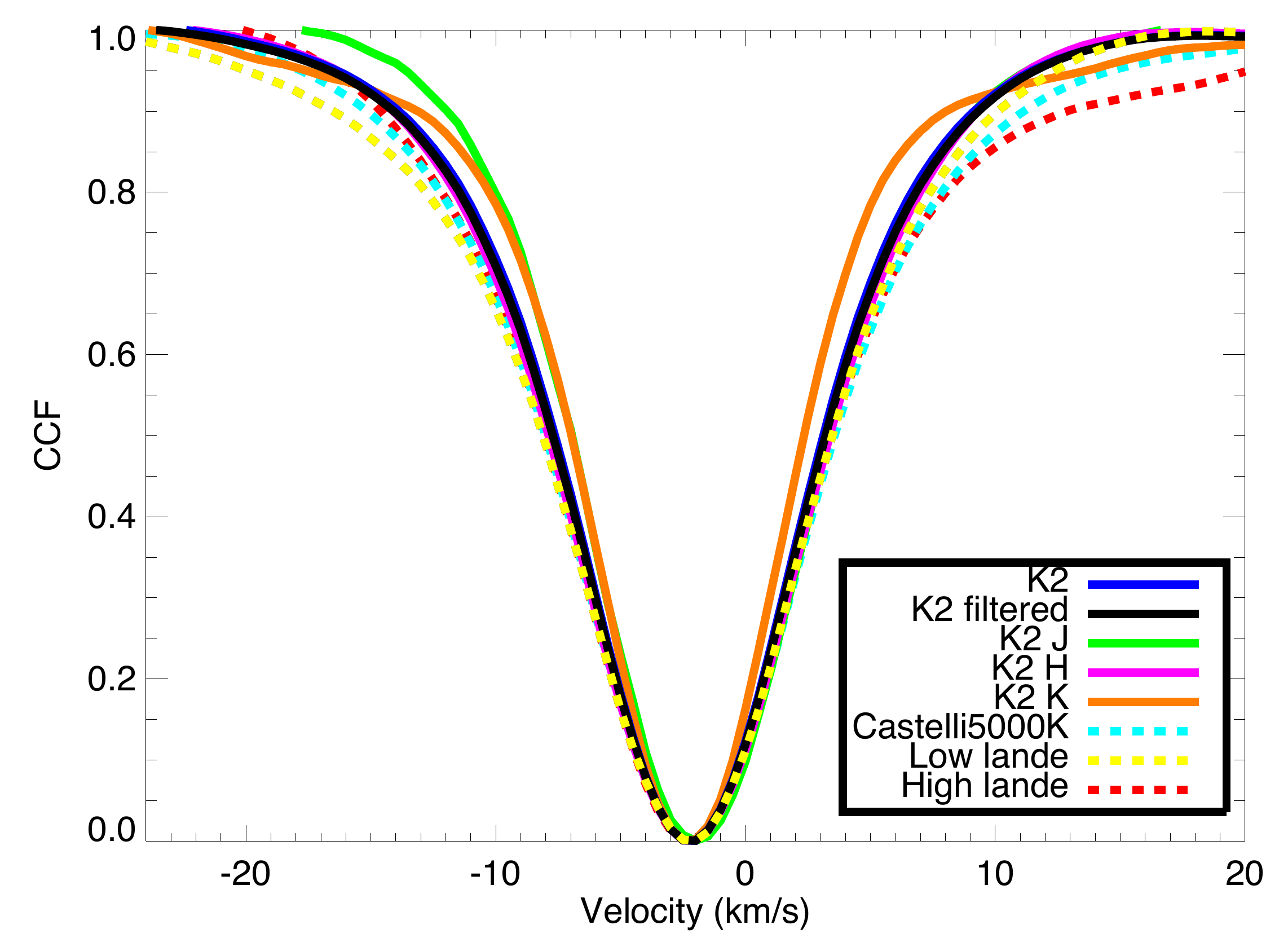}
   \includegraphics[width=0.9\hsize]{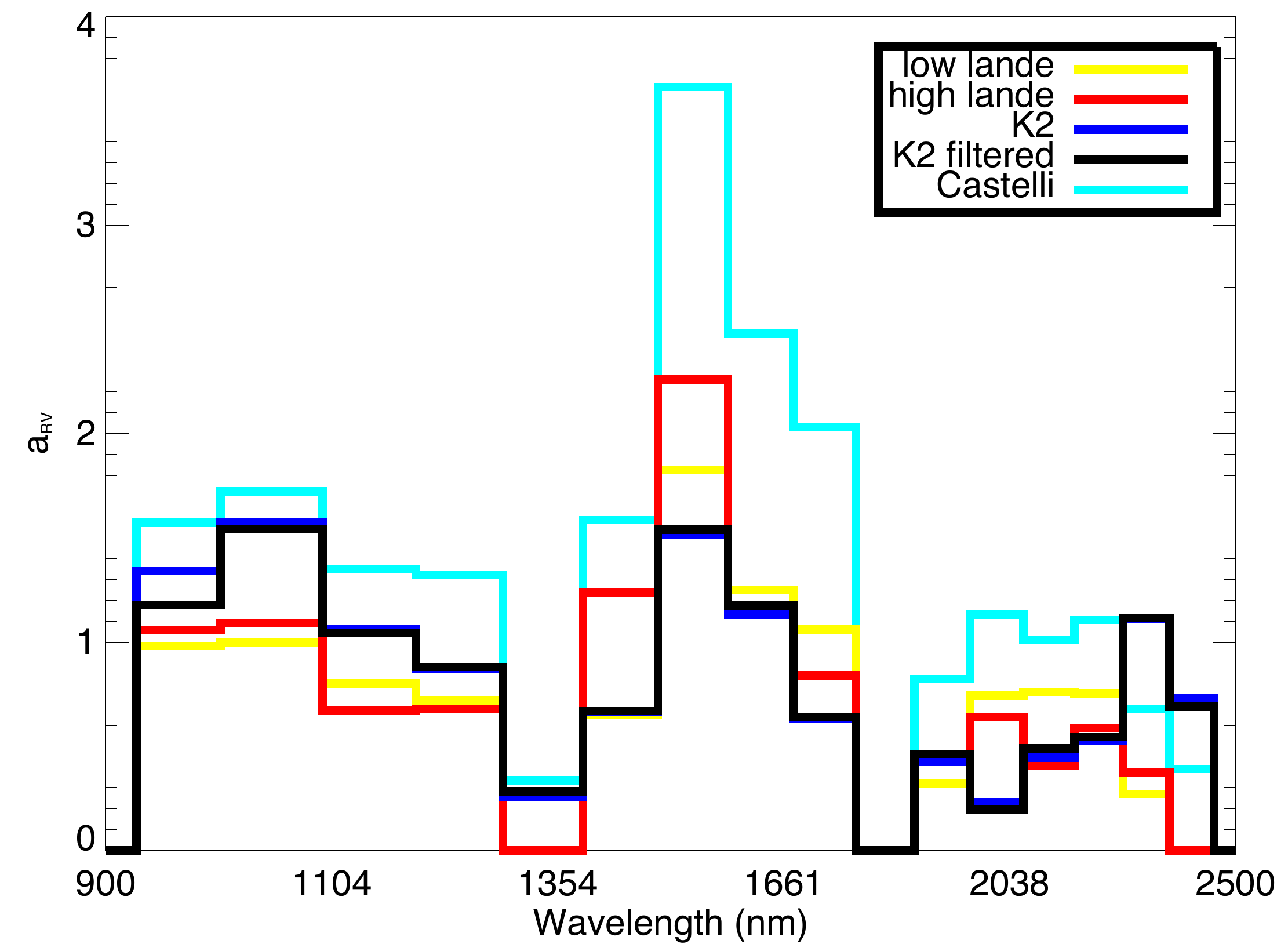}
      \caption{(Top) Cross-correlation functions obtained with \spirou\ pipeline APERO using a variety of numerical masks: K2, an empirical mask based on the whole collection of spectra, its separate $J$, $H$, and $K$ contributions and a more filtered version of it, and a theoretical mask based on Castelli models containing atomic lines (with its two flavors of low-Land\'e factor and high-Land\'e-factor lines, see text). (Bottom) Distribution of $a_{RV}$ showing the relative contribution of each wavelength bin to the RV precision for different line lists.
              }
         \label{masks}
   \end{figure}

Figure \ref{masks} (top) compares the average CCF obtained with each mask. Table \ref{table:2} shows the number of lines, the average FWHM and spectral shifts obtained on the cross-correlation functions with these masks. The number of lines is given in two ways: the smaller number is the number of lines in the line list, and the larger number in parenthesis includes all duplicate occurrences of lines when consecutive orders overlap (hence the larger difference in the $J$-band mask, where the overlaps are wider).

It might be expected that the $K$-band lines are either of similar width or slightly wider than the $J$-band lines because the Zeeman-splitting behavior increases with wavelength squared. We observe a 10\% wider $H$-band CCF compared to the $J$-band CCF. 
However, the $K$-band CCF in turn is 17\% narrower than the H CCF. 
The $K$ band includes a larger fraction of molecular lines, which have a narrower profile than atomic lines, and the average profiles might reflect this as the dominant factor. Because these CCFs are not built from similar line densities, this discrepancy may also be due to a lower precision of the $K$-band CCF because the FWHM has a scatter four times larger for the $K$ band than the $H$ band. 
In addition, the comparison between the low-Land\'e and high-Land\'e masks shows no systematic broadening for the second one, within a dispersion of 0.2 km/s. This requires further investigations using other stars with various activity levels. The comparison of CCF width between the theoretical mask (Castelli) and the empirical mask (K2) shows that the second is 1 km/s narrower than the first. This may indicate that some reference wavelengths in the theoretical line list are slightly misplaced.

Finally, the K2-$J$ line list shows a spectral shift compared to the reference K2 mask. If this shift were further investigated, it might show that some lines are contaminated by telluric residuals or residual blended stellar lines.

The RV precision of each wavelength bin is proportional to $a_{RV} =  \sqrt N \times <I> $, where $<I>$ is the average intensity of the lines and $N$ is the number of lines in the bin. 
The bottom plot of Figure \ref{masks} shows that the distribution of $a_{RV}$ over the spectral domain is quite different between the theoretical and empirical lists: while it is similar in the two sets of masks, it is twice lower in the $H$ and $K$ bands in the empirical mask than in the theoretical mask. The low- and high-Land\'e masks have a similar wavelength distribution.

      \begin{table*}
      \caption[]{Characteristics of the average CCF for a variety of line lists (see text). Column 2 lists the real number of stellar lines in the mask and the total number of line occurrences when order overlaps are included. Columns 3 and 4 give the average FWHM of the CCF and the velocity shift compared to the K2 CCF; errors represent their scatter in the sample. The last two columns list the $rms$ values of the residuals to the orbit model and the classical Rossiter-MacLaughlin model.}
         \label{table:2}
         \begin{center}
         \begin{tabular}{p{0.2\linewidth}llcccc}
            \hline
            \noalign{\smallskip}
  mask   &  nb lines & FWHM   & $\delta v$&$rms$ orbit & $rms$ RM \\
         &          & km/s    & km/s &  m/s  & m/s \\
            \noalign{\smallskip}
            \hline
            \noalign{\smallskip}
Castelli           &4121 (6288) &12.6$\pm$0.1  &  0.06$\pm$0.08&11.6 & 29.0\\
low Land\'e          &1323 (2034) & 12.7$\pm$0.08  & -0.11$\pm$0.06&47.2 & 22.2 \\
high Land\'e         &1289 (1987) &12.6$\pm$0.2 & 0.04$\pm$0.11 &40.5& 17.4\\
K2                 &1924 (3246) &11.5$\pm$0.1  &0.00$\pm$0.07 &10.8 & 5.9 \\
K2 filtered      & 1405 (2378)& 11.7$\pm$0.1 & 0.002$\pm$0.07 &10.2 & 5.3\\
K2  $J$              &646 (1895) &10.7$\pm$0.1  & 0.46$\pm$0.08 &41.1 & 26.0\\
K2  $H$              &894 (959) &11.9$\pm$0.1  &0.004$\pm$0.07&12.2 & 9.1\\
K2  $K$              &384 (392) & 9.9$\pm$0.4  &-0.07$\pm$0.07& 24.7 & 17.0\\
            \noalign{\smallskip}
            \hline
         \end{tabular}
         \end{center}
   \end{table*}

\subsection{Dispersion and conditions\label{dispersion_and_conditions}}
In this section, we compare photon noise RV estimates and observed RV scatter in this data set featuring the K star \hd189.

The pipeline estimates the RV uncertainty of each individual exposure using the optimal weight method on the spectrum, as introduced by  \citet{connes1985} and \citet{bouchy2001}. Because this is performed on the extracted spectrum rather than the CCF, this RV uncertainty indicator does not depend on the mask and underestimates the real value because telluric lines are still included. Moreover, while the photon-noise estimator in the pipeline gives relevant values for most spectra, some spectra had erroneous values that did not scale with S/N. For these, we estimated the photon-noise uncertainty by using the scaling observed in the majority of spectra. The scaling function is \begin{equation}
    \delta RV_{rms} = 240 / S/N + 0.65 m/s
    \end{equation}
    where S/N is the S/N per 2.28-km/s pixel at 1700 nm (note that this relation can only be used for \spirou\ observations of a star whose spectral type and rotational velocity are similar to  those of \hd189). The chosen threshold for correction was $\pm$10\% of the expected value. After this correction, the average RV precision of the data set is 1.8 m/s. Individual nightly averages of the estimated photon-noise RV uncertainty are given in Table 1 in the penultimate column, RV phn.

As each circular-polarisation sequence consists of four individual spectra, it is possible to extract and derive the spectrum position of all subexposures as opposed to the position of the combined Stokes I spectrum, and derive an estimate of the RV scatter inside a sequence, although small-number statistics applies here. 
The last column, RV $rms$, in Table \ref{table:1} shows the nightly RV dispersion values. The lowest dispersion values obtained in this data set and with version 0.5.0 of the data reduction system are 3-4 m/s, or about twice the average photon noise. This reaches higher values when the extinction varies or the S/N is low. Worse seeing conditions in turn do not lead to significantly higher dispersion values in this limited data set. 

\section{Radial-velocity analysis\label{radial_velocity_analysis}}
\subsection{Orbit}
\begin{figure}
   \centering
   \includegraphics[width=\hsize]{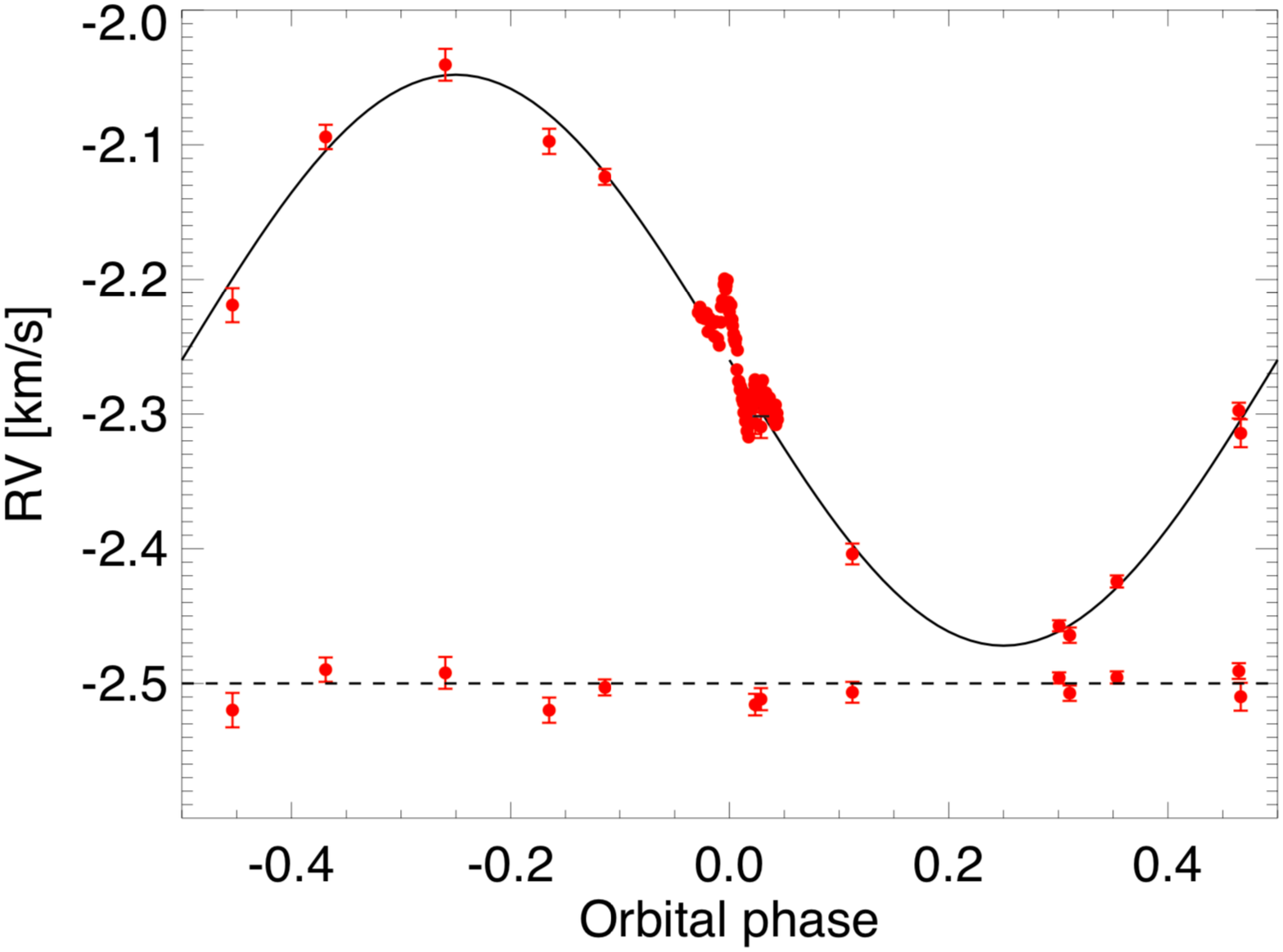}
      \caption{Average RV per polarimetric sequence taken over the orbital cycles of the planetary system 
      as a function of orbital phase with residuals to the literature model. The data in this plot are listed in Table \ref{tab:rvs}.
              }
         \label{FigOrbit}
   \end{figure}

All RVs as measured with the K2-filtered mask are listed in Table \ref{tab:rvs}.
The radial velocity measurements obtained with \spirou\ were then phased according to the following ephemeris \citep{baluev2019}: 
\begin{equation}
\phi_{\rm orb}=(BJD-2458383.8012)/2.21857545
\end{equation}
As shown in  Figure \ref{FigOrbit}, the recovery of the planetary orbit as known from the literature is secured from this sparse data set, although the time series is spread over three commissioning runs from July to October 2018, and various changes were made in the instrument between runs (both hardware and software). It shows that daily calibrations operate as expected, by deriving the correct velocity shift from spectral lamp observations. The hardware changes during the six-month commissioning period were not considered significant enough to justify adding a free offset between runs in this study, because there were only a few visits.

The $rms$ of these O-C residuals is 10.8\,m/s rms with the K2 cross-correlation mask, when combined RVs per polarimetric sequences are used. The value is 10.2\,m/s with the K2-filtered mask. This residual scatter is compatible with the $rms$ jitter previously found by \citet{bouchy2005} (15\,m/s, ELODIE), \citet{winn2006} (12\,m/s, HIRES), and \citet{boisse2009} (9\,m/s, SOPHIE), and with the 33.5\,m/s peak-to-peak variations reported by \citet{triaud09} between several transit sequences (HARPS). These were obtained with optical spectrographs.

We obtain a dispersion of O-C residuals of: 41.1\,m/s with the K2 mask limited  $J$, 12.2\,m/s in  $H$, and 24.7 in  $K$. 
The dispersion is much smaller in $H$ within a polarimetric sequence and over the full time series. 

\subsection{Rossiter-McLaughlin effect}
\label{data_rm}

The two transit sequences obtained in spectroscopic mode can be used to measure the Rossiter-McLaughlin effect. We can compare this with previous results from optical measurements. The K2-filtered mask was used for this analysis. 
The improvement in RV precision with this mask is about 1\,m/s compared to the K2 mask, that is, an improvement of about 20\%, but more importantly it removes a residual systematic slope of about 0.07\,km/s/day, which is probably due to residuals from telluric lines that slowly move over the stellar spectrum with Earth rotation during the sequences.

\subsubsection{Classical Rossiter-McLaughlin fit}
\label{RM_sec}
The Rossiter-McLaughlin effect was fit using the model developed by \citet{ohta2005}. This model derives an accurate analytic formula for the radial velocity anomaly and takes the stellar limb darkening into account. In order to determine the limb-darkening coefficient ($u$) for \hd189 b in the infrared range, we used 3D stellar model atmospheres from \citet{hayek2012}. A simple linear limb-darkening law was used to derive an approximate value of $u = 0.433$ for the $J$ band.

Five free parameters were used to simultaneously fit the two transits of \hd189 b: the sky-projected obliquity $\lambda$, the projected rotation velocity $v\sin i_{\star}$, the transit epoch $\tau$, and two systematic velocities $\gamma_{1}$ and  $\gamma_{2}$. All other transit and Keplerian parameters were fixed at the values reported in the literature (see Table \ref{table:rm}).

\begin{figure}
   \centering
   \includegraphics[width=\hsize]{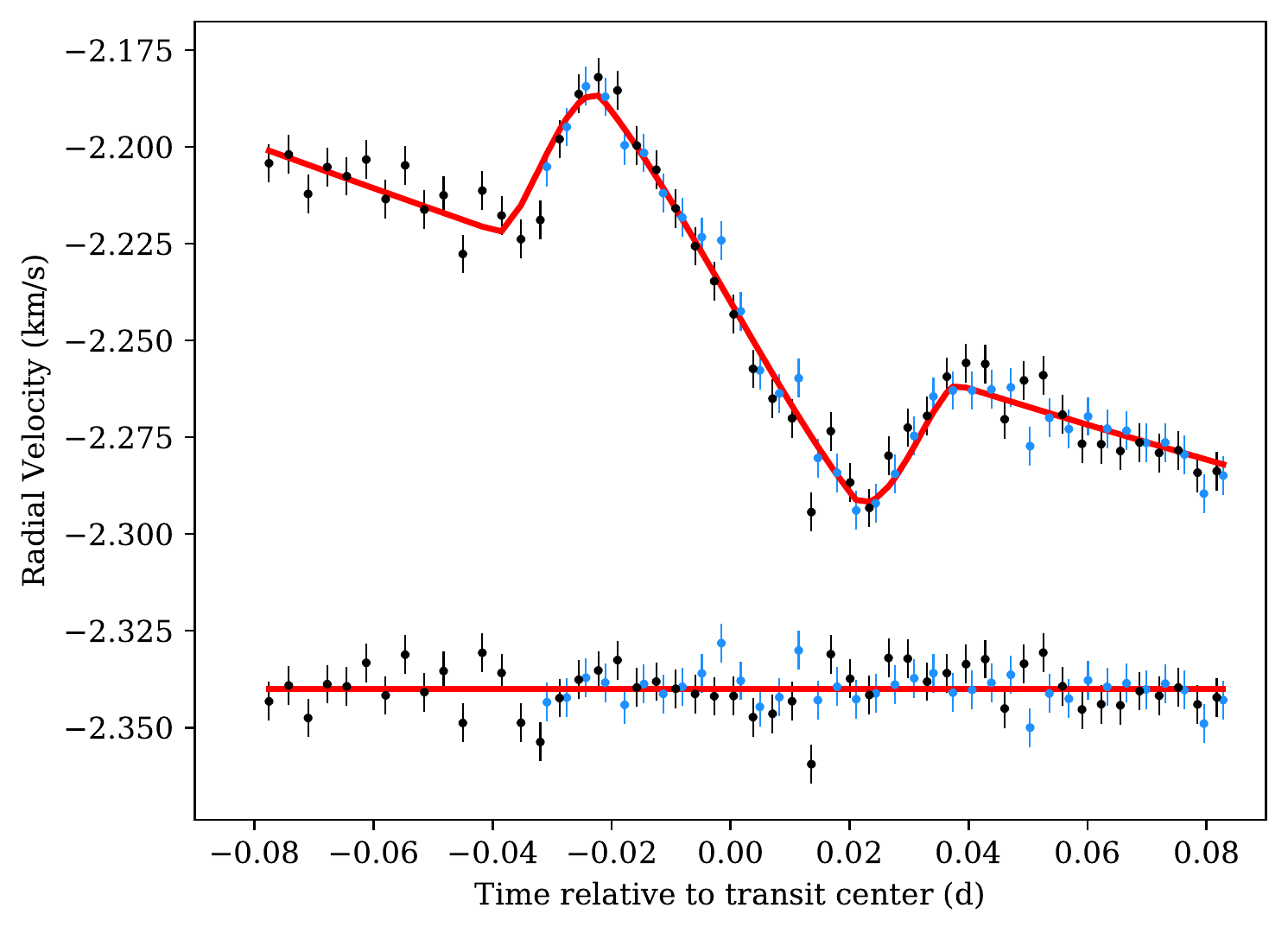}
    \caption{Individual RV measurements of the two observed transits as obtained from the CCF analysis (top; blue: first sequence, black: second sequence), together with the best-fit model obtained with the MCMC procedure (solid red line); (bottom) residuals after subtracting the best-fit model are shifted by -2.34 km/s for better visualization.
              }
         \label{FigRM}
   \end{figure}

We sampled the posterior distributions of $\lambda$, $v\sin i_{\star}$, and $\tau$ using the Markov chain Monte Carlo (MCMC) software \texttt{emcee} \citep{foreman2013}, assuming uniform priors. We used 50 walkers and 2000 steps of the MCMC and discarded the first 500 steps. Best-fit values are the medians of the distributions, with 1~$\sigma$ uncertainties that are derived by taking limits at 34\% on either side of the median, as listed in Table \ref{table:rm}. Figure \ref{FigRM} shows the RV measurements of \hd189\ taken during the two transits of planet b on September 21, 2018, and June 14, 2019, along with the best-fit model. We found a sky-projected obliquity $\lambda$ = $ -3.6\stackrel{+1.5}{_{- 1.4}}$ degrees and vsini$_{\star}$ = $3.29\stackrel{+0.09}{_{-0.09}}$  kms$^{-1}$. When we fit this separately, the second sequence has a smaller offset on the $\lambda$ value than the reference optical value, $-1.9 \pm 1.8$ degrees, compared to when the two sequences were combined, probably because the first sequence is not complete. It starts slightly after ingress and has no baseline before the transit. These partial sequences understandably lead to increased systematics. 
These values obtained with one or two \spirou\ transits agree within 2$\sigma$ with the values obtained from data in the optical range: $\lambda$ = $ -0.85\stackrel{+0.32}{_{-0.28}}$ degrees and vsini$_{\star}$ = $ 3.316\stackrel{+0.017}{_{-0.067}}$ kms$^{-1}$ from four transit sequences with HARPS, as analyzed by \citet{triaud09} or revised later with $\lambda$ = $-0.4\pm0.2$ degrees from \citet{cegla2016}.

\begin{figure}
   \centering
   \includegraphics[width=\hsize]{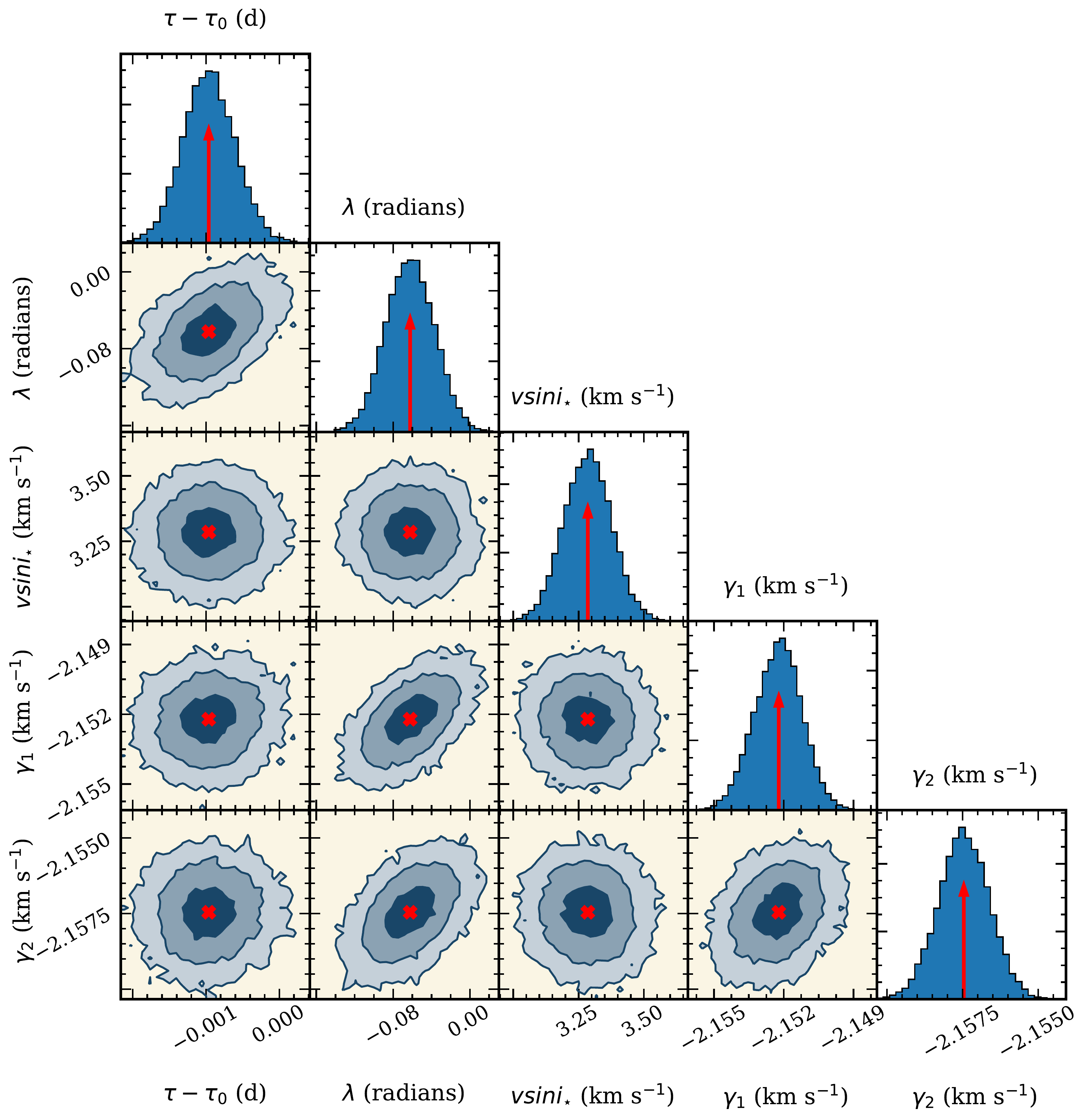}
      \caption{ Corner plot of the Rossiter-McLaughlin effect when the two transit sequences are combined. Different color counters marks the 1$\sigma$, 2$\sigma$, and 3$\sigma$ regions. The red arrows in the histograms correspond to the distribution medians.
              }
         \label{cornerplot}
   \end{figure}

The dispersion of residuals after the subtraction of the model is 5.31\,m/s for both transits, or 4.08\,m/s and 6.04\,m/s for the September 2018 and June 2019 transit, respectively. Out-of-transit RV data have a dispersion of 3.82 and 5.25\,m/s for the first and second transit sequences (15 and 27 baseline data points), respectively, after the orbital slope is removed. This number likely represents the current instrumental scatter that was measured on early-science \spirou\ data of a bright K star with this version of the pipeline. For comparison, when the full K2 mask is used or when the measurement is limited to the $J$, $H$, and $K$ bands, the dispersion of the residuals for both transits are 5.9, 26, 9, and 17\,m/s, respectively, when the two transits are fit together, with the residual slope mentioned earlier.  The final fit parameters did not change significantly for different values of the limb-darkening coefficient $u$ corresponding to each band ($u_{Y} = 0.467$, $u_{H} = 0.343$ and $u_{K} = 0.286$).  In Figure\,\ref{cornerplot}, we show the correlation diagrams for all five free parameters. 
The best-fit value of the transit epoch, $\tau$, is consistent with the value of ($\tau_{0}$) in \citet{baluev2019} by propagating over different orbits using the orbital period from Table\,2 of \citet{baluev2019}.

 \begin{table*}[t]
\caption[]{Best-fit parameters of the Rossiter-McLaughlin effect. }
\label{table:rm}
\begin{tabular}{l c c c c}
Parameter & Unit & Prior type & Value & Reference\\ 
\hline 

Period  & days& Fixed & 2.21857545 & \citet{baluev2019} \\ 
$K$  & kms$^{-1}$ & Fixed & 0.201 & \citet{boisse2009} \\ 
$\omega$  & degrees & Fixed & 90.0 & \citet{winn2007} \\ 
$e$  &  & Fixed & 0.0028 & \citet{baluev2019} \\ 
$a/R_{\star}$  &  & Fixed & 8.7566 & \citet{triaud09} \\ 
$i_{\star}$  & degrees & Fixed & 85.712 & \citet{baluev2019}\\ 
$r/R_{\star}$  &  & Fixed & 0.15703 & \citet{baluev2019} \\ 

\hline 
Fit Parameters\\ 
&  &  &  \\
$\tau$  & BJD & $\mathcal{U}$ (2458383.800,2458383.803)    &   $2458383.8012189\stackrel{+0.00036}{_{-0.00035}}$ & This work \\
$\lambda$ & degrees &  $\mathcal{U}$ ($-10$,$10$)& $ -3.6\stackrel{+1.5}{_{-1.4}}$ & This work  \\
v$\sin i_{\star}$ & kms$^{-1}$ & $\mathcal{U}$  (2,5)    & $3.29\stackrel{+0.09}{_{-0.09}}$ & This work\\
 $\gamma_{1}$ & kms$^{-1}$  &$\mathcal{N}$  ($-2.3$,0.1)  & $-2.15221\stackrel{+0.001}{_{-0.001}}$  & This work\\  
 $\gamma_{2}$  & kms$^{-1}$ & $\mathcal{N}$  ($-2.3$,0.1) & $-2.15745\stackrel{+0.0008}{_{-0.0008}}$  & This work \\
             \noalign{\smallskip}
            \hline

\end{tabular}
\tablefoot{
$\mathcal{U}$(min,max) corresponds to uniform probability distribution with min and max as the minimum and maximum values and $\mathcal{N}$(mean,sigma) corresponds to Normal probability distribution.}
\end{table*}

 \subsubsection{Doppler tomography}
 
\begin{figure}
   \centering
   \includegraphics[width=\hsize]{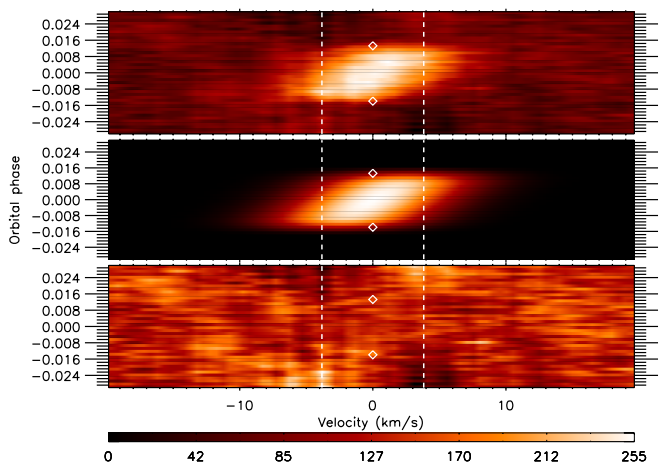}
    \caption{ Doppler Tomography results obtained on the June 2019 transit: the diamonds mark the ingress and egress from literature ephemeris and the dashed line corresponds to the v$\sin i_{\star}$ of the star. Top panel: residuals after subtracting the model stellar CCF from the CCFs obtained. Middle panel: best-fit model corresponding to $\lambda = -0.5\degree$. Bottom panel: residuals after subtracting the best-fit model.
              }
         \label{DT}
   \end{figure}

We applied another technique, Doppler tomography, to the same data to determine the obliquity measurement for \hd189 b and checked its consistency
with the classical RM method. The 50 CCFs from the second transit were used for this analysis. We considered a model of stellar CCF that includes a limb-darkened rotation profile convolved with a Gaussian corresponding to the intrinsic photospheric line profile and instrumental broadening, following the approach of \citet{cameron2010}. The Gaussian bump due to the planet occultation was also modeled in the stellar line profile, whose spectral location depends on the planet position in front of the stellar disk during the transit. We made a tomographic model that depends on the same parameters as the classical RM fit (Sect. \ref{RM_sec}), and added the local line profile width, $s$ (nonrotating local CCF width) expressed in units of the projected stellar rotational
velocity \citep{cameron2010,dalal2019}. The free parameters we used to fit the Gaussian bump were $\lambda$, v$\sin i_{\star}$, $\gamma$ and $s$. The $\tau$ was fixed to $\tau_{0}$ and the other parameters were taken from the literature. The top panel in Figure \ref{DT} shows the bright signal corresponding to the stellar surface regions that are occulted by \hd189 b.  The best-fit model as shown in the middle panel of the Figure \ref{DT} was obtained with $\lambda$ = $ -0.5\stackrel{+1.2}{_{-1.3}}$ degrees and v$\sin i_{\star}$ = $ 3.80\stackrel{+0.16}{_{-0.16}}$ km/s. The obliquity and rotational velocity from the classical RM and the Doppler tomography models agree well (See Sect. \ref{RM_sec}). 

This is the first time that two different techniques, a classical RM fit and Doppler tomography, were used to measure obliquity in the NIR range. They are both
compatible with the results obtained from the optical data. The data from the second sequence modeled alone gave a result closer to the optical data, although with slightly larger error bars; this is explained by the incomplete transit sequence of the first epoch, although it was observed in better conditions than the second epoch.

\section{Stellar features \label{stellar_features}}

\subsection{Activity jitter}

\begin{figure}
   \centering
   \includegraphics[width=0.9\hsize]{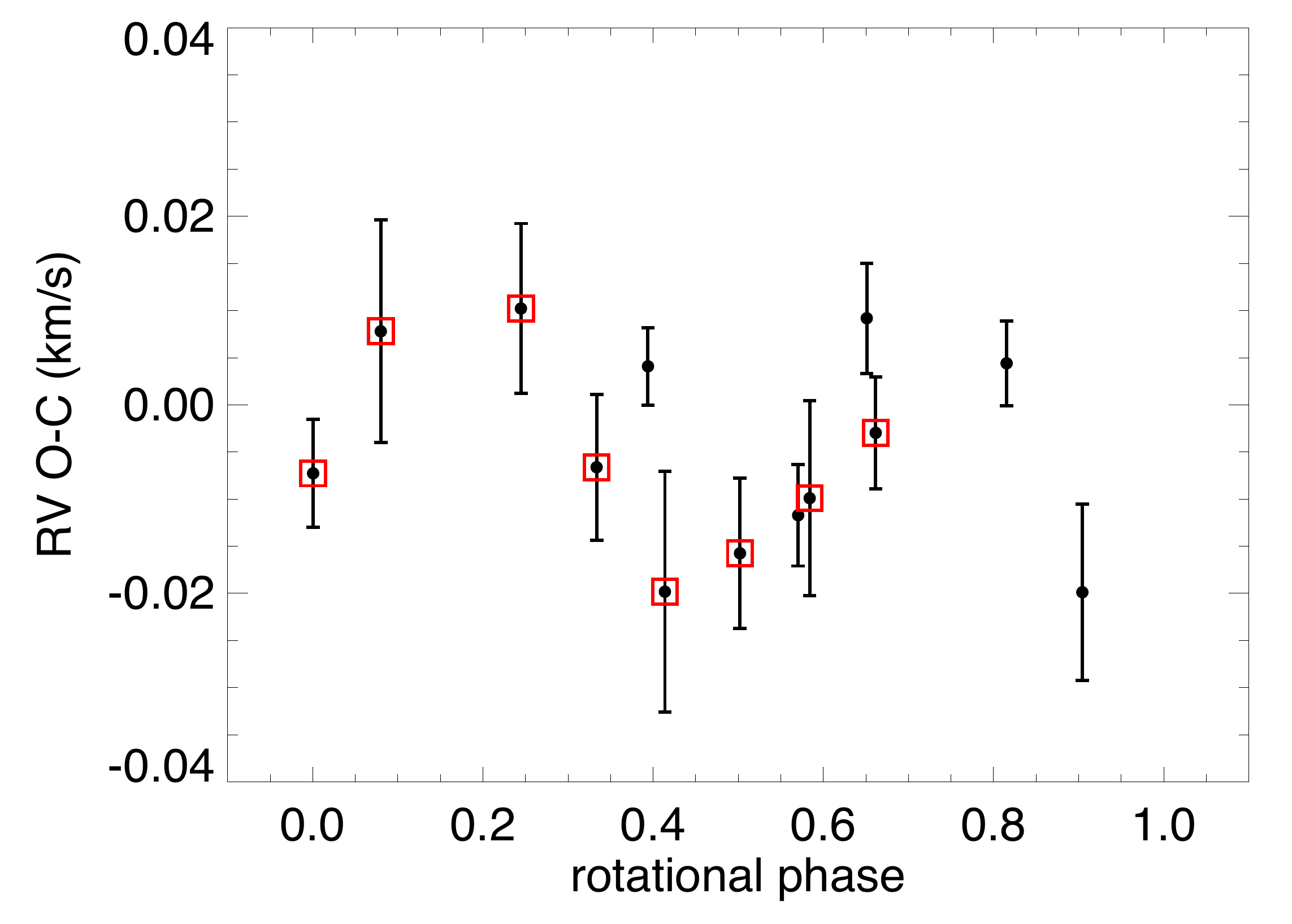}
      \caption{Residual RV jitter as a function of the rotational phase of the parent star. Red squares indicate the first rotational cycle.
              }
         \label{rotOC}
   \end{figure}

\hd189\ is known to have an average rotation period close to 12 days \citep{henry2008} while  a differential rotation of 0.11\,rad/d has been measured by Zeeman Doppler imaging \citep{fares10,fares17}. We used an average period of 12 days with the following ephemeris: 
\begin{equation}
\phi_{\rm rot} =(BJD-2458329.018)/12.0
\end{equation}
Figure \ref{rotOC} shows the behavior of the residual RVs from the planetary orbit subtraction as a function of the rotational phase of the parent star.  As earlier, the error bars reflect the relative scatter of the four consecutive RV measurements that were made within a polarimetric series (last column of Table 1). The residual RVs are quite dispersed in this data set, which extends over three months (eight rotational cycles). The peak-to-peak variations are about 30\,m/s.  Considering the eight first visits, which were obtained within a dozen days, that is, during a single rotational cycle, a more organized pattern appears with a maximum near phase 0.1 and a minimum near phase 0.4 before another rise (points with a red square in Figure\,\ref{rotOC}). This suggests that  cool spots cluster around rotation phase 0.3 at the time of our first eight observations. The rapid evolution of the stellar photosphere is well known \citep{boisse2009,lanza2011}, and it is not surprising that the five last RVs do not seem in phase with the first RVs, as they are separated by 47\,days from the first batch, and spread over another 34\,days. The data set is not sufficient for a relevant jitter analysis or for attempting to correct for it, but the possible dependence of the residuals on rotational phase may indicate that at least part of the RV residuals is due to stellar activity.  

The FWHM of the CCF averaged over each polarimetric sequences does not correlate with the RV residuals of the orbit (Pearson coefficient -0.08). It may differ from observations made in the optical where the FWHM seems to be a good indicator of the stellar activity when it is due to a spot or a plage \citep{dumusque2014a}. This difference might also indicate that in \spirou\ data, the CCF broadening is dominated by Zeeman broadening rather than spot and plage temperature contrast, and hence is less prone to rotational modulation. Alternatively, residual systematic noise in the intensity profile might affect its shape. Even in the more frequently studied cases of optical CCFs, the origin of variations in the FWHM is not always clear and may be more sensitive to instrumental systematics, as shown earlier for SOPHIE data of \hd189\ \citep{dumusque2014b} and other cases observed with HARPS-N \citep{benatti,mayo2019}.

   \begin{figure}
   \centering
  \includegraphics[width=0.9\hsize]{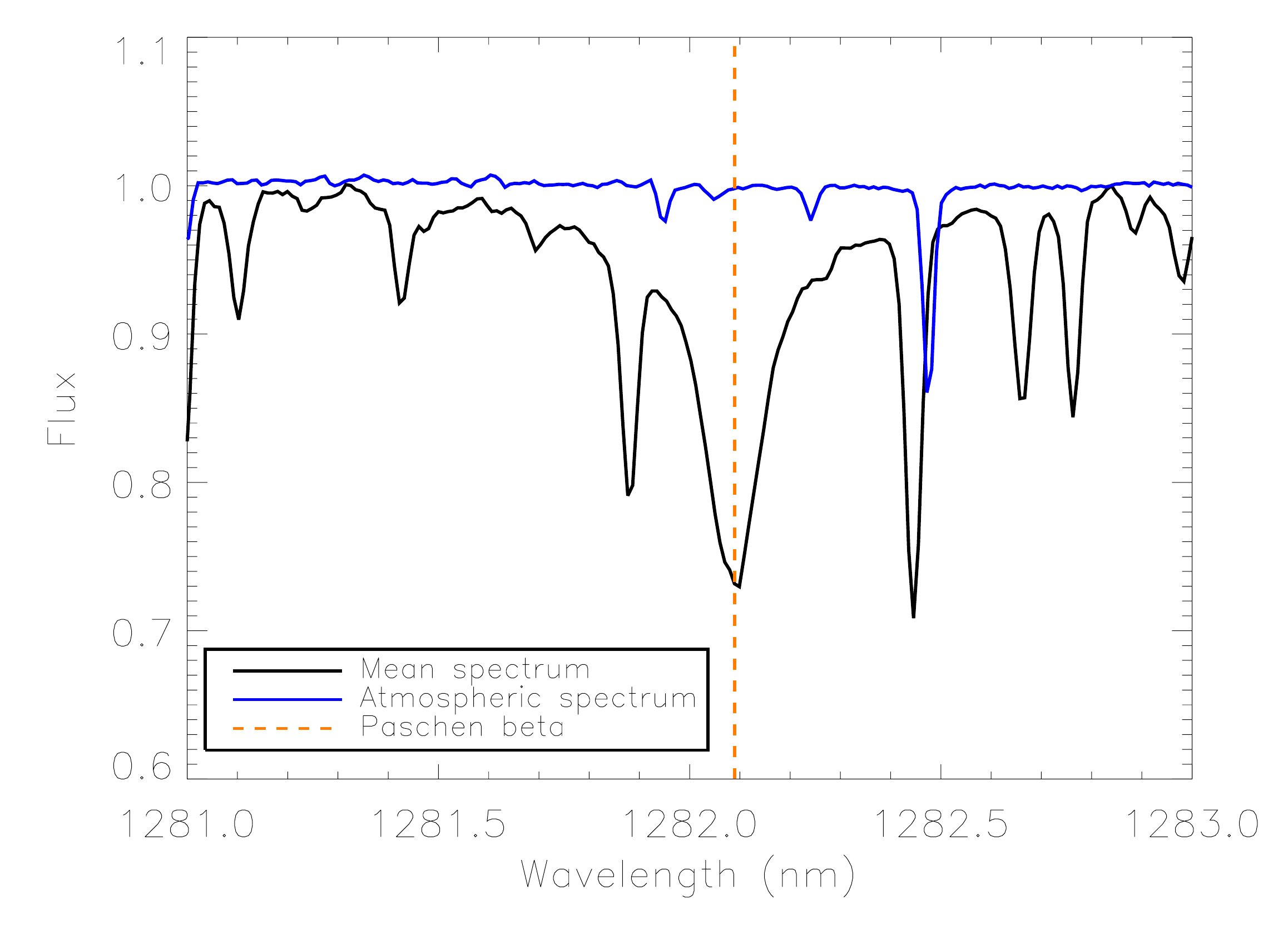}
  \includegraphics[width=0.9\hsize]{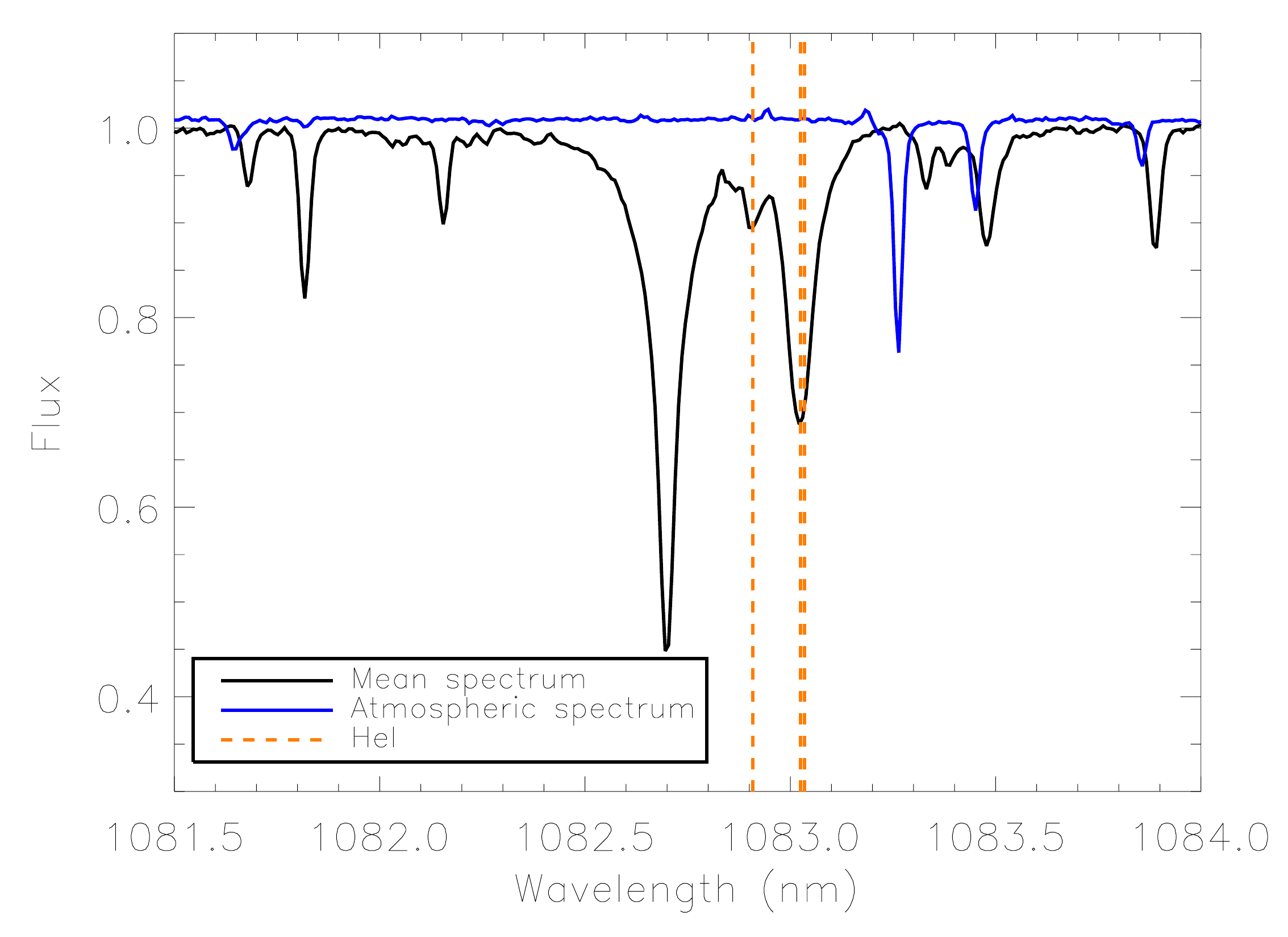}
      \caption{(Top) Mean telluric-corrected spectrum of \hd189 (in black) near the Pa$\beta$ line (vertical line). The subtracted telluric spectrum is shown in blue. (Bottom) Same for the HeI triplet line. 
              }
         \label{FigActivity}
   \end{figure}

\subsection{Pa$\beta$ and HeI activity indices}
In order to compare the RV jitter to an independent activity criterion, we searched the spectra for chromospheric lines that would be sensitive to stellar modulated phenomena. We selected the Paschen $\beta$ line at 1282\,nm and the HeI triplet at 1083\,nm. The Pa$\beta$ spectral region is shown in Figure \ref{FigActivity} (top) together with the reconstructed spectrum of telluric contamination (a byproduct of the telluric correction), which has very little effect in this domain. Two methods were compared. In the first, a Gaussian fit was adjusted to the core of the line in all spectra as well as to the mean spectrum. An average equivalent width per polarimetric sequence was then calculated after the mean value was removed. In the second method, the mean spectrum was first removed from each spectrum and the residual in the stellar line was integrated over the line width. The error of each visit was calculated to be the scatter within the polarimetric sequence. The two methods give similar trends, and the second method has slightly less noise than the first and is considered in the analysis. The telluric-subtracted spectra are used. 

As a sanity check, this method was used on other lines close to the HeI and Pa$\beta$ regions. Three lines at 1084.680, 1282.520, and 1283.495 nm were then measured in the same way as HeI and Pa$\beta$. The first two are contaminated by telluric residuals at some values of the BERV, and the last line is not contaminated. This comparison allowed us to set a threshold on any significant variability, accounting for the variable noise in the spectra. The main conclusion is that there is no significant difference between the variable behavior of the HeI and Pa$\beta$ lines with respect to the check lines. 

\subsection{Spectropolarimetry}

The Stokes V Zeeman signatures for our \spirou\ observations were calculated using the ratio method as described in \cite{donati97} and \cite{bagnulo2009}, where we used a sequence of four exposures consisting of two pairs of observations that each correspond to different retarder position. For each exposure in the sequence we extracted the two orthogonal-polarization science channels individually and then applied the blaze flux correction.  
Then a wavelength correction was applied to each exposure to compensate for the BERV. Finally, the flux values of each exposure were interpolated using a B-spline to match the wavelength sampling of the first exposure in the sequence. Then we calculated the degree of polarization for each spectral point using the ratio method.  We measured the pseudo-continuum of the polarization spectrum using a moving median with a window width of about 1000 points. Then we fit a low-order polynomial to the measured continuum points to model the continuum in the whole spectrum, which was further subtracted from the data.

\begin{figure}
   \centering
    \includegraphics[width=0.9\hsize]{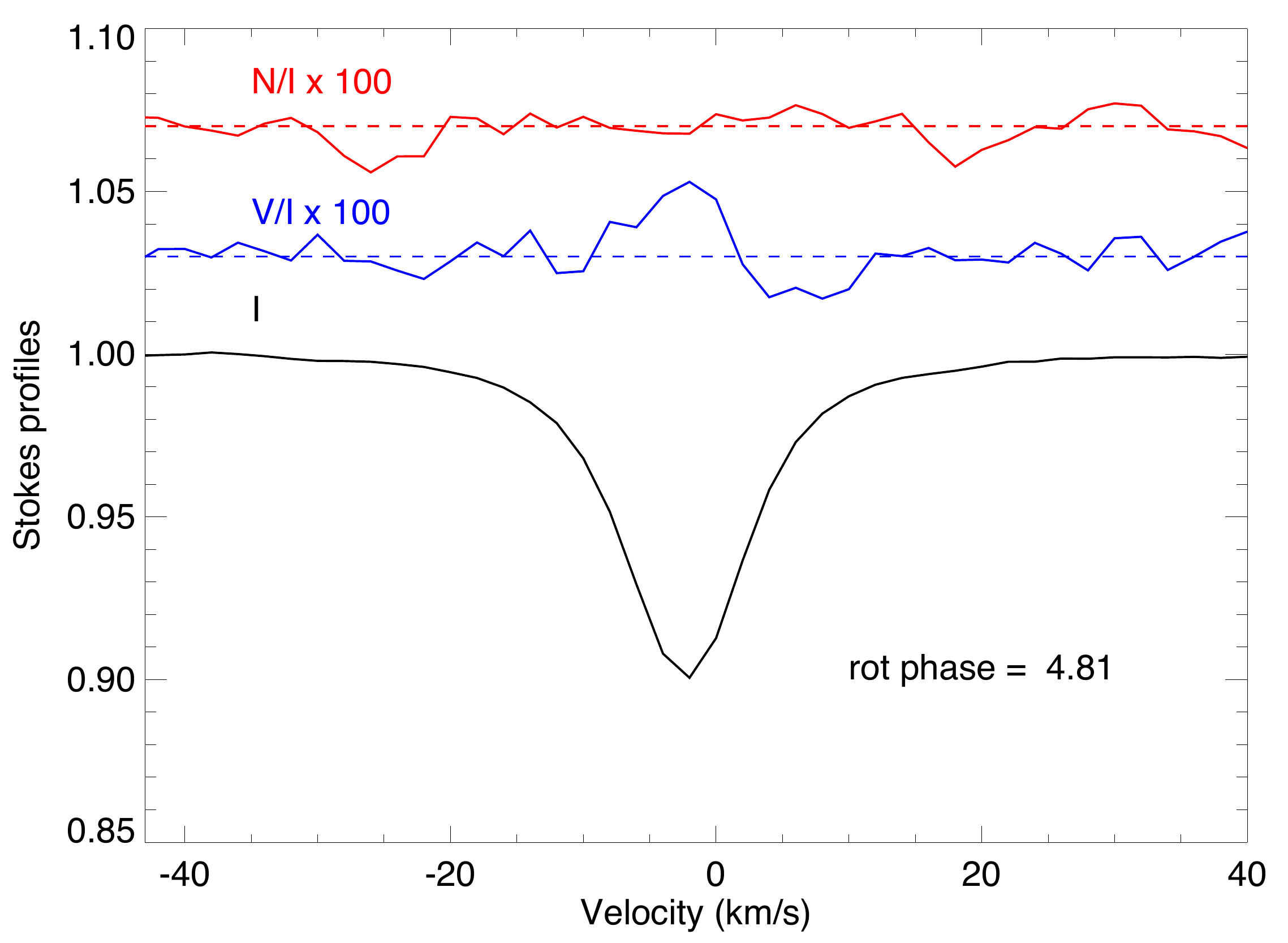}
    \includegraphics[width=0.9\hsize]{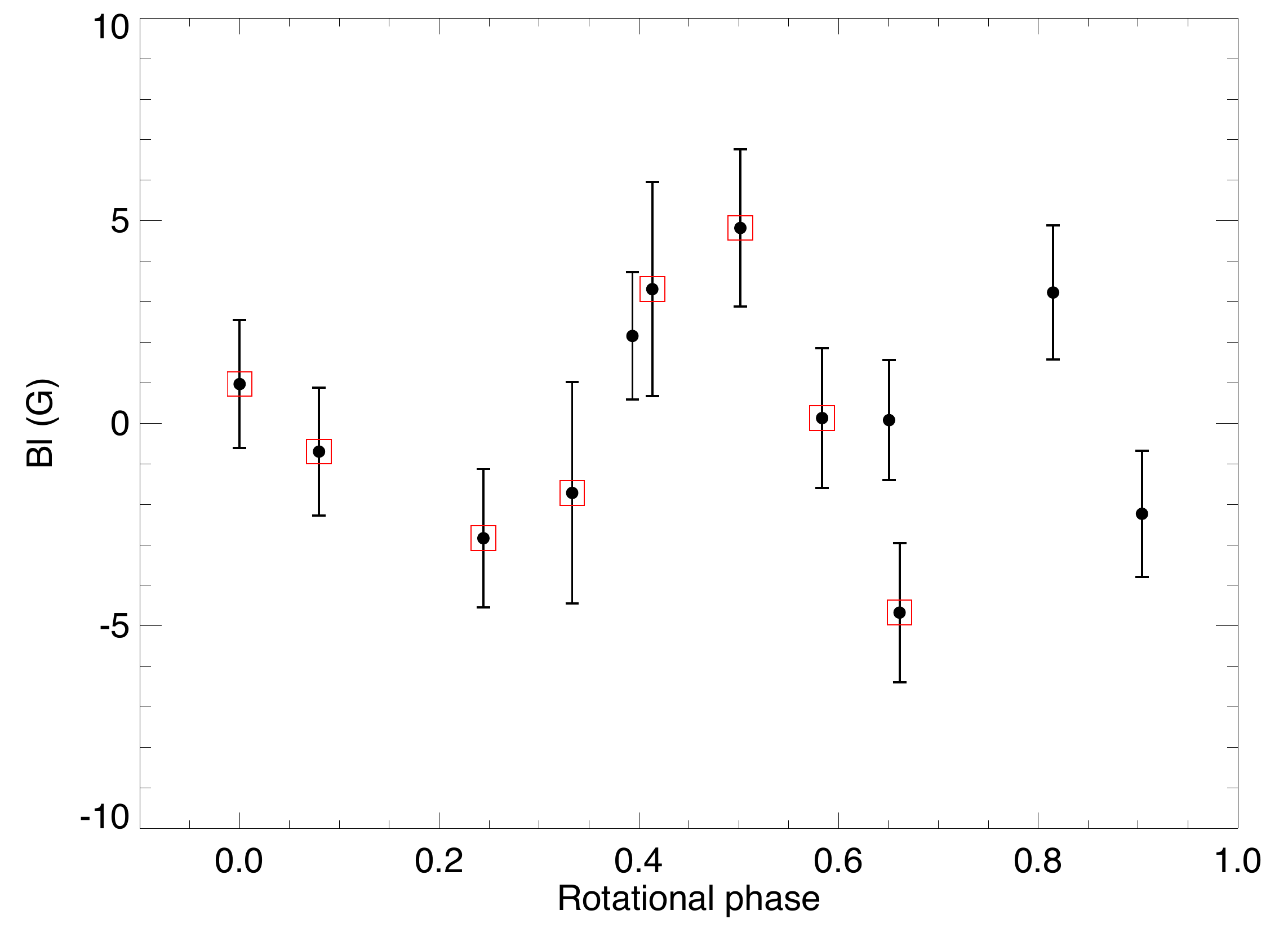}
      \caption{(Top) LSD Stokes profiles for one of the polarimetric sequences showing a definite detection in the V profile. The Stokes V and N profiles are multiplied by 100 and shifted for clarity. (Bottom) The variations in the longitudinal magnetic field as a function of the rotational phase. Red squares indicate the first rotation cycle.
              }
         \label{FigLSD}
   \end{figure}

In order to obtain the circular polarization (Stokes V) in the stellar lines, we used the least-square deconvolution technique (LSD)  \citep{donati97} with the Castelli mask that contains the Land\'e factor of each line, that is, the sensitivity of each line to the stellar magnetic field. These factors are currently only available through physical experiment or theory and are lacking for most molecular lines in the NIR spectrum. The exclusion parameter we used to remove spectral domains that are contaminated by telluric residuals corresponds to telluric lines that are deeper than 5\%. Between the Land\'e factor restriction and the cut for telluric residuals, only 1366 lines were used in the polarization profile. An example of Stokes V signatures is shown in Figure~\ref{FigLSD}. It also shows the LSD profile of the null spectrum (N), obtained from a different combination of individual spectra and aimed at confirming the potential instrumental artifacts \citep{donati97}.

Table~\ref{table:lsd} lists all polarimetric sequences and their characteristics: stellar rotational phase, maximum signal and standard deviation in the Stokes V profile, and longitudinal field B$_\ell$. The latter was obtained from the following equation:
\begin{equation}
    B_\ell=-2.14 \times 10^{11} \frac{\int vV(v)\mathrm{d}v}{\lambda_0 \cdot g_{\rm eff} \cdot c \cdot \int \left[ I_c - I(v) \right] \mathrm{d}v}
\end{equation}
where the mean wavelength $\lambda_0$ is 1512 nm and the average Land\'e factor $g_{\rm eff} $ is 1.25. The values for B$_\ell$ span a few Gauss for the few profiles in which a signature is detected in the Stokes V profile, in accordance with the values reported in the optical with ESPaDOnS \citep{moutou07,fares10,fares17}. The median error on B$_\ell$ is 1.7 Gauss, which is to be compared with a median error of 1 Gauss in optical literature values, while exposure times that were used with ESPaDOnS are two to three times longer. It is expected that with similar exposure times, the error on the longitudinal field would have been similar between ESPaDOnS and \spirou. Moreover, the number of available lines with a known Land\'e factor is larger in the wavelength domain of NARVAL (0.37--1.0 $\mu$m) than in \spirou\ (0.98-2.35 $\mu$m) for a K dwarf star.
The Zeeman sensitivity, however, is higher in the NIR than the optical, which more or less compensates the lower number of lines. This explains the similar error bars on B$_\ell$ for comparable exposure times.

Correcting for telluric contamination before calculating the LSD profiles does not significantly change the Stokes V profile (at a level less than 1$\sigma$), which is expected because the ratio method acts as a filter against any signatures that do not change in a short time or with the polarization state, including telluric features.
As seen previously, the theoretical mask is not optimal for deriving the radial velocity at high precision because it does not include molecular lines, in contrast to the empirical mask. 
Further iterations between the theoretical line lists and observed spectra are required to improve the Stokes V profiles, for instance, by including strong molecular features with known Land\'e factors. 

Figure \ref{FigLSD} (bottom) shows the variation in the longitudinal magnetic field with respect to the stellar rotation phase. As for previous indicators, the variation is smooth during the first rotational cycle, while additional visits acquired several cycles later show no clear behavior. Although the longitudinal field and the RV residuals show indications of rotationally modulated evolution, these two quantities do not appear to be directly correlated with each other in this data set, similarly as has been observed for other stars \citep{hebrard2016,hussain2016}. Furthermore, the data collection is not sufficient to try modeling the magnetic topology of the stellar surface with the few detected signatures spread over several months.

      \begin{table*}
      \caption[]{Characteristics of least-squares deconvolution Stokes V profiles obtained for all polarimetric sequences. $B_\ell$ is the longitudinal magnetic field in Gauss. $Bf$ is the magnetic modulus calculated from the Zeeman broadening of FeI lines (section \ref{zbsection}).}
         \label{table:lsd}
         \begin{center}
         \begin{tabular}{llccccc}
            \hline
            \noalign{\smallskip}
  Date     &  Rot. & Max signal  & $\sigma$ & $B_\ell$ &$Bf$\\
  -2458000 & phase & 10$^{-4} I_c$ & 10$^{-4} I_c$  & (G) & (G)\\
            \noalign{\smallskip}
            \hline
            \noalign{\smallskip}
 329.018&    0.00&    2.17&    0.35&    0.97$\pm$  1.58 &271$\pm$68\\
 329.973&    0.08&    2.16&    0.37&   -0.70$\pm$  1.57 &267$\pm$48\\
 331.949&    0.24&    2.30&    0.43&   -2.84$\pm$  1.71 &236$\pm$75\\
 333.016&    0.33&    1.77&    0.51&   -1.72$\pm$  2.73 &342$\pm$65\\
 333.972&    0.41&    1.67&    1.48&   -0.31$\pm$  2.75 &324$\pm$74\\
 333.986&    0.41&    1.39&    0.92&    6.93$\pm$  2.53 &357$\pm$80\\
 335.038&    0.50&    2.94&    0.58&    4.82$\pm$  1.94 &250$\pm$76\\
 336.020&    0.58&    2.15&    0.35&    0.13$\pm$  1.72 &282$\pm$50\\
 336.951&    0.66&    2.00&    0.39&   -4.67$\pm$  1.72 &183$\pm$46\\
 384.825&    4.65&    0.21&    0.41&    0.08$\pm$  1.48 &332$\pm$36\\
 386.797&    4.81&    2.26&    0.41&    3.23$\pm$  1.65 &332$\pm$43\\
 387.866&    4.90&    1.72&    0.56&   -2.23$\pm$  1.56 &302$\pm$68\\
 417.740&    7.39&    2.62&    0.45&    2.16$\pm$  1.57 &262$\pm$49\\
            \hline
\end{tabular}
\end{center}
\end{table*}

\subsection{Zeeman broadening}
\label{zbsection}
Zeeman broadening provides an alternative diagnostic of the stellar magnetic field, because it infers a magnetic field strength from line widths in the Stokes I spectrum.  This provides a measure of magnetic field strengths on small scales, which are invisible in Stokes V because regions with opposite sign cancel each other out \citep{morin2013}. Because Zeeman broadening scales with wavelength squared while most other broadening processes scale with wavelength, the ability to detect this effect is much higher with \spirou\ than with optical spectrographs.  

The approach taken here was to directly model carefully chosen spectral lines with a wide range of effective Land\'e factors and minimum blending.  We used the Fe {\sc i} lines at 1534.3788, 1538.1960, 1561.1145, and 1564.8510 nm.  The 1564.8510 nm line was used by \citet{Valenti1995} and \citet{lavail2017}. Table \ref{table:ZBlines} lists the line characteristics used in this study. The choice for these specific lines is the result of a trade-off analysis using the following criteria: large (and, respectively, very small) Land\'e factors, large line depths, no blending, and long wavelengths. Atomic and molecular lines in the $K$ band that are used in M stars \citep[e.g.,][]{flores2019}, for instance, are too weak in the spectrum of \hd189.

\begin{table}
\caption[]{Atomic data of the Fe {\sc i} lines used in the spectral synthesis including Zeeman broadening. $g_{\rm eff}$ is the effective Land\'e factor, $\log gf$ the oscillator strength, and $\log C_6$ is the van der Waals damping coefficient. $\log gf$ for Fe {\sc i} 1564.85 is from  \citet{Valenti1995}, other $\log gf$ values are from fits to the solar spectrum by Petit et al.\ (2020, in prep.), and other data are from \citet{Kurucz-K14}. 
}
         \label{table:ZBlines}
         \begin{center}
         \begin{tabular}{llccc}
            \hline
            \noalign{\smallskip}
 $\lambda$ (nm) & Element & $g_{\rm eff}$ &$\log gf$& $\log C_6$ \\
            \noalign{\smallskip}
            \hline
1534.3788 & Fe {\sc i}& 2.63& -0.67& -7.52\\
1538.1960 & Fe {\sc i}& 0.01& -0.69& -7.43\\
1561.1145 & Fe {\sc i}& 1.83& -3.30& -7.79\\
1564.7413 & Fe {\sc i}& 1.00& -1.08& -7.29\\
1564.8510 & Fe {\sc i}& 2.98& -0.63& -7.49\\
\noalign{\smallskip}
            \hline
\end{tabular}
\end{center}
\end{table}

Synthetic spectra including Zeeman splitting were calculated using the {\sc Zeeman} spectrum synthesis code \citep{Landstreet1988, Wade2001, Folsom2016}.  This produces local-thermal-equilibrium spectra in all four Stokes parameters, although here we focused on Stokes I.  A grid of MARCS model atmospheres \citep{Gustafsson2008-MARCS-grid} was used as input.  While {\sc Zeeman} accurately reproduces atomic lines in LTE for K stars \citep[e.g.,][]{Folsom2016}, it does not calculate molecular line spectra. We therefore avoided molecular lines.

Atomic data for four Fe {\sc i} lines were taken from the VALD database \citep{Kupka1999-VALD, Ryabchikova2015-VALD3}, but the oscillator strengths were not sufficiently accurate for this analysis.  For the Fe {\sc i} 1564.8510 nm line we adopted the empirical $\log gf$ of \citet{Valenti1995}.  For the other three lines we fit the oscillator strengths using an observation of the solar spectrum.  The observation was obtained with \spirou\ in sunlight reflected off the moon.  Synthetic spectra were calculated for a solar model atmosphere and oscillator strength $\log gf$ values were fit through a $\chi^2$ minimization routine.

Spectral synthesis was performed using the stellar effective temperature (4875 K) and gravity (4.56) from interformetric observations \citep{boyajian2015}. A v$\sin i_\star$ value of 3.47 km/s was inferred from the equatorial rotation period and inclination of the rotation axis from \citet{fares17} and the stellar radius of \citet{boyajian2015}. We assumed a microturbulence of 0.9 km/s, which is typical for this spectral type \citep[e.g.,][]{Doyle2013}. These parameters were considered as fixed. Then the spectral fit derived the following parameters: macroturbulence, metallicity, a magnetic field strength, and a filling factor for this magnetic field.  A radial-tangential macroturbulence profile was used.  The magnetic field was assumed to have a uniform radial geometry, covering a fraction of the surface given by the filling factor.  This is clearly simpler than the real magnetic geometry, but Zeeman broadening is relatively insensitive to magnetic geometry, and this is the simplest geometry consistent with the observations, and is consistent with models used in previous Zeeman broadening studies \citep[e.g.,][]{Valenti1995}.

\begin{figure*}
   \centering
    \includegraphics[width=0.7\hsize, angle=270]{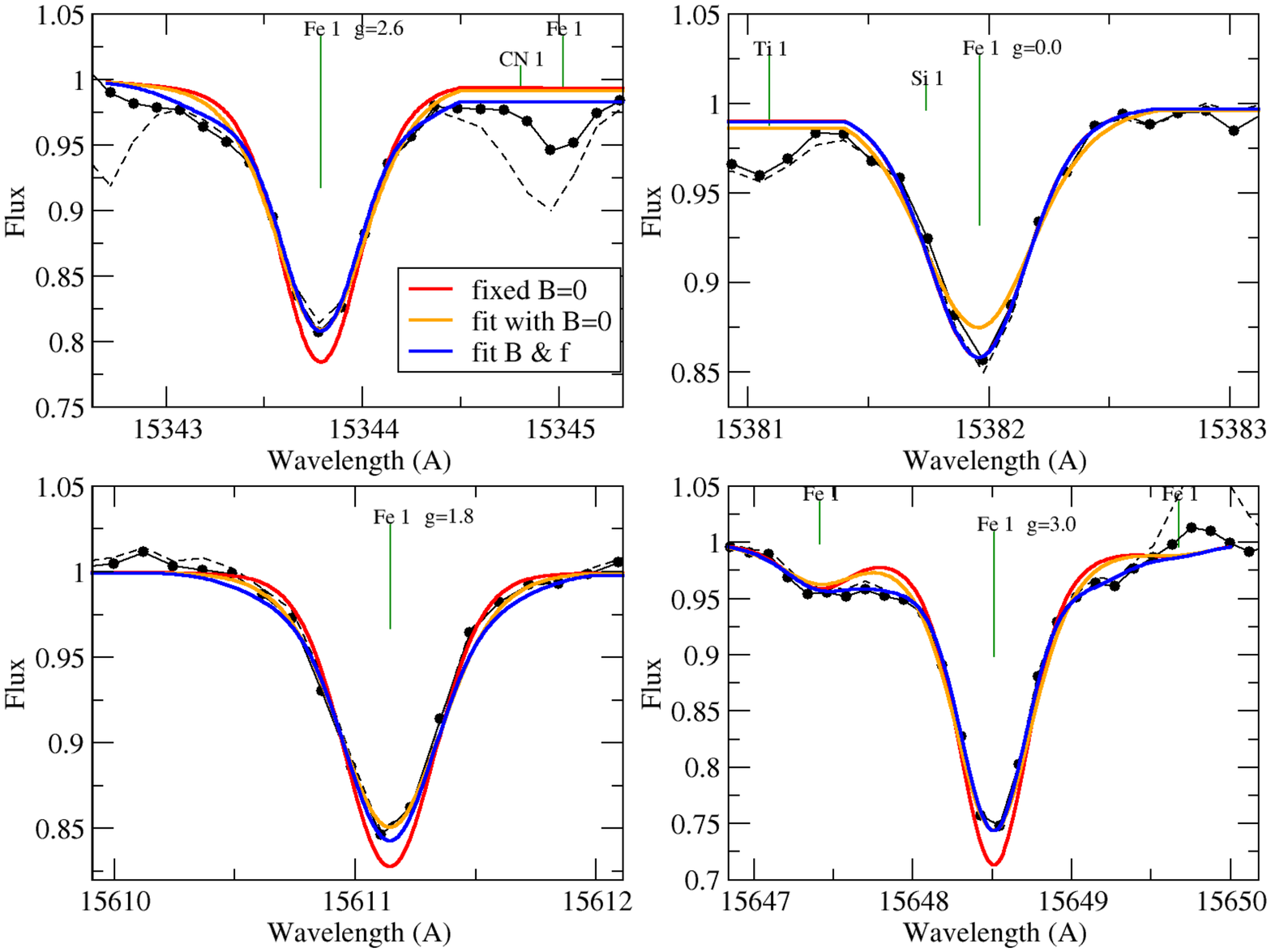}
      \caption{Observed (black line and dots) and modeled spectra of \hd189 in four lines of Fe I with various Land\'e factors (see the $g$ value in the line label). Three models are compared: a model spectrum with fixed spectroscopic parameters and a magnetic field fixed to zero (red line); a model with fitted spectroscopic parameters in which all lines contributing equally to the fit, even the broadened ones and no magnetic field (orange line), and a model in which all parameters are fit (blue line) (see text). The dashed line represents the spectrum that is not corrected for the telluric lines.
              }
         \label{FigZB}
   \end{figure*}

The four telluric-subtracted spectra were coadded for the visit with the highest S/N (September 23) and were then generalized to all epochs. 
A plot of the fitted lines for September 23 is shown in Figure \ref{FigZB} together with a comparison of the best fit of the four lines with a null magnetic field. This modeling shows that a nonzero magnetic field is required to fit the wings of the higher Land\'e factor lines, while the model also fits the width of the zero Land\'e factor line. The red line shows the line profiles when spectroscopic parameters are fixed and without Zeeman broadening: all three lines with significant $g$ appear to be deeper. Finally, the best-fit model with additional magnetic flux parameters is shown as the blue line and matches all line profiles. 

\begin{figure}
   \centering
 \includegraphics[width=0.9\hsize]{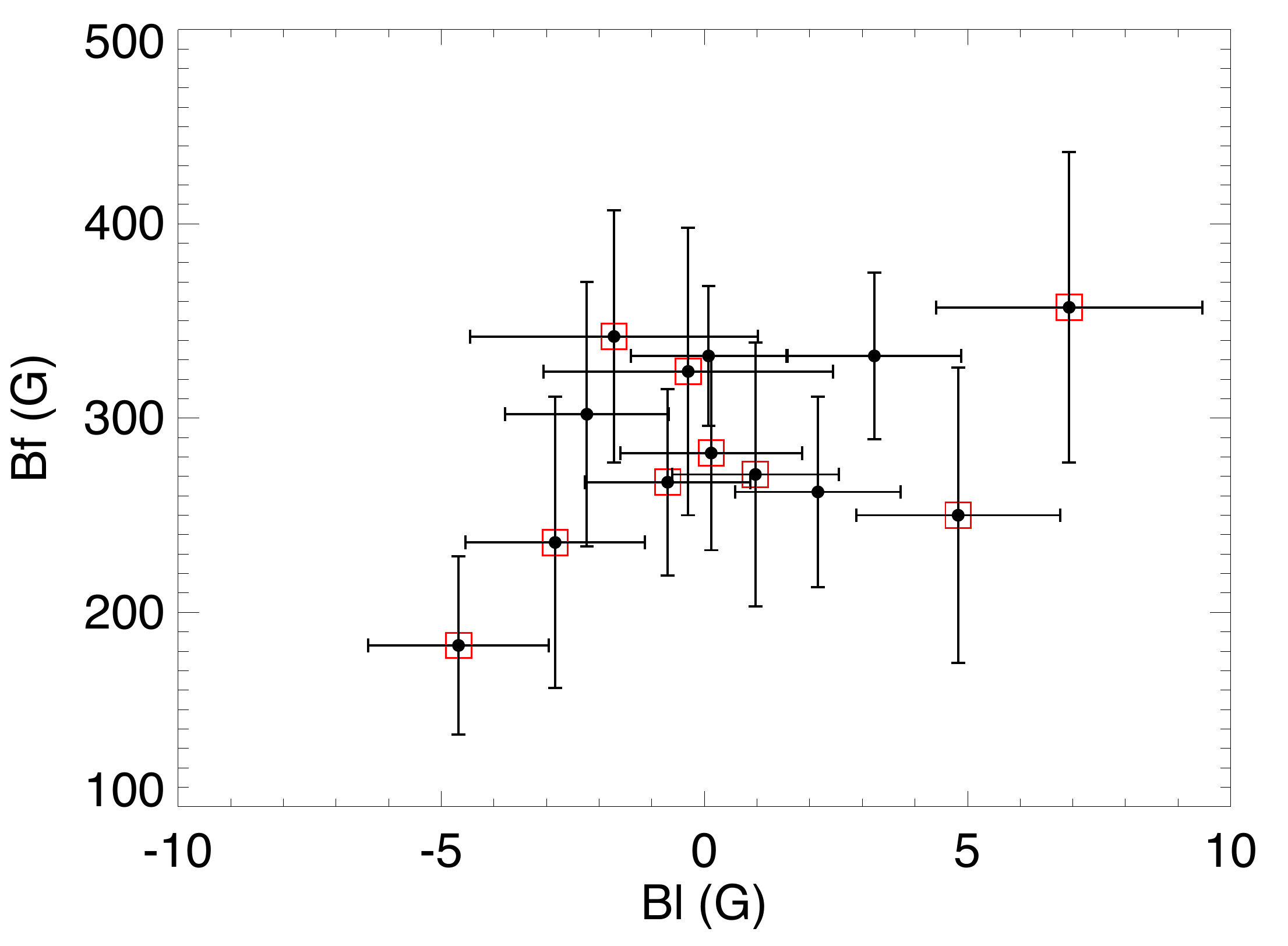}
      \caption{Magnetic field flux $Bf$ as a function of the longitudinal  field  $B_\ell$. Red squares indicate the measurements made during the first rotational cycle.          }
         \label{BlBf}
   \end{figure}

The best-fit macroturbulence is 3.5$\pm$0.3 km/s and metallicity of 0.09$\pm$0.02 (in 1.3$\sigma$ agreement with the literature value of -0.03$\pm$0.08 \citep{torres2008}). On average for all epochs, \hd189\ has a magnetic field of 1.9$\pm$0.2 kG that covers 15$\pm$2\% of the star; the average magnetic flux $Bf$ is 290$\pm$58 G. Individual uncertainties are obtained from the covariance matrix, scaled by the square root of the reduced $\chi^2$. A few sources of errors are not included (on parameters that were considered as fixed: Teff, gravity, and microturbulence), therefore these uncertainties may be slightly underestimated. The last column in Table \ref{table:lsd} lists the individual $Bf$ values corresponding to each polarimetric sequence. While there is a mild correlation between the longitudinal field and the field flux, as shown in Figure \ref{BlBf}, the latter cannot be related to RV residuals. 

This value of the magnetic field strength and flux for \hd189 is intermediate between nonactive, slow-rotators M dwarfs such as Gl~411 or Gl~380, which have typical fields lower than 1 kG \citep{moutou2017,flores2019}, and fast-rotator M dwarfs whose field usually ranges from 2 to 5 kG \citep{shulyak2019}. With a rotational period of 12 days, \hd189 is similar to Gl~49, DS~Leo, Gl~208 or CE Boo, whose magnetic field fluxes lie between 0.5 and 2 kG according to far-red measurements of FeH and Ti (see Figure 5 of \citet{shulyak2019}). Comparisons of the filling factor of 15\% we determined with the filling factor of other solar-type dwarfs agree well with data in the compilations by \citet{cranmer2011} and \citet{see2019}, as this parameter is highly variable. For a Rossby number of 0.403 as estimated for \hd189 \citep{vidotto2014}, Figure 4 in \citet{see2019} shows a range of magnetic filling factors of 5 to 50\%.

\section{Discussion and conclusion}

We observed the planet-hosting star \hd189\ with \spirou\ during its science verification phase and early operations. The \spirou\ data allowed the recovery of the orbital signature and spectropolarimetric features in agreement with previous work done at optical wavelengths. 

From the flux in the spectra of \hd189, we can estimate the photon-noise RV uncertainty in the total NIR spectrum to be about 1.8\,m/s for an S/N of about 250, using the method described in \citet{bouchy2001}.  A more comprehensive study of the photon-noise distribution in \spirou\ spectra was not included in the objectives of this article, and will be the object of future studies. In particular, it would be interesting to study whether the photon noise or the instrumental or processing scatter (or which portion of each) is responsible for the observed difference between the relative $J$, $H$, and $K$ observed RV dispersion values. The scatter in the $K$ band RV data is currently twice larger than in the $H$ band (Table 2), in line with the relative weight of these bands shown in Figure 2. However, the situation is less clear for the $YJ$ bands: while their contribution to the RV precision is expected to be as large as (or larger than) the $H$ band, the observed scatter is about three times larger. Moreover, the CCF calculated with the $YJ$ lines alone shows a redshift compared to the CCF obtained with both the $H$ and $K$ bands. Whether this is due to a differential effect of stellar activity or to a stronger effect of uncorrected absorption or emission lines in the $YJ$ bands remains to be investigated.
Furthermore, a comparison with similar analyses based on RV content of spectra of various spectral domains and resolving power \citep{figueira2016,cloutier2018,reiners2020} would help in understanding the specific limitations of \spirou\ data for this type of stars. 

The observed RV scatter over a timescale of an hour has a median of 6\,m/s and it is $\sim$10\,m/s over a timescale of a few months. If the short-term scatter represents the combination of photon-noise contribution and instrumental or processing jitter, then the activity jitter of \hd189 in the period July to October 2018 is about 8\,m/s $rms$. There is an indication of rotation modulation in this additional scatter. Because \hd189 is known to be a star with an active surface, a comparison with the RV jitter that is usually attributed to stellar activity from optical data cannot be definite at this stage: it would require contemporaneous observations. A residual activity jitter of about 8\,m/s, however, is plausible, given the range of values from 9 to 15\,m/s of the literature values obtained in the optical domain (see Section 3.1 and references therein). While observing the stellar surface in the NIR range might damp the jitter because the spot contrast is lower than in the optical \citep{reiners2010}, it might also enhance it with larger Zeeman broadening \citep{reiners2013}. A precise comparison would therefore clearly require simultaneous campaigns with NIR and optical spectrographs that would ideally be equivalently equipped with spectropolarimetry and high-precision velocimetry for a precise diagnostics.

Further ways to improve the instrumental or processing RV precision budget are being investigated. One way is to increase the data set with a larger BERV span, in order to improve the stellar template. Overall telluric correction will also become more robust with the use of additional principal components, as allowed by a larger data set. Another way is to use template matching (as in \citet{astudillo2015}, e.g.) when this better-quality template is available. With more adjusted telluric corrections, the line list can also be enlarged with a net improvement expected on the RV precision as well as on the retrieval of activity indices such as bisector span and FWHM of the CCF. A more accurate wavelength calibration is also being implemented in the \spirou\ pipeline. Secondary effects, such as contamination from the calibration channel or persistence from previous observations, are finally being investigated. 
A main difference between NIR and optical RV data sets is indeed that spectra are not independent of each other. Not only is the telluric subtraction dependent on the relative velocities, but flux contamination may occur between channels or in time (detector persistence, and smearing or electronic crosstalk from the calibration channel). While these effects are not expected to dominate the error budget for \hd189, they should nevertheless be estimated when these corrections will be implemented and validated by testing. The data set presented in this paper is particularly adapted to test further improvements of the pipeline, when available, because the achromatic signatures of the planet are well known.  

For the observation of two transits of \hd189\, it was possible to model the Rossiter-McLaughlin signal in two different ways (direct modeling and Doppler tomography), which both gave results that agree well with the optical analysis from the literature. The Doppler tomography result from the second, complete, transit, gives a projected obliquity of $-0.5\pm1.3$ degrees, which agrees very well with the modeling of four transit sequences observed with HARPS, which allowed measuring $-0.4\pm0.2$ degrees \citep{cegla2016}. The spectroscopic analysis of these transits in search for planet atmospheric signatures is ongoing and will be published in a forthcoming paper.

The fast-evolving activity of the stellar surface of \hd189 has been studied and modeled multiple times in the past \citep[e.g.,][]{ boisse2009,lanza2011,fares17} because in many ways, it affects the measurement of the system parameters, including atmospheric composition and planetary environment conditions \citep{pont2007,sing2011,bourrier2020}. The stellar magnetic field of \hd189 was also analyzed in the \spirou\ data set. From the intensity spectrum, we were able to fit the unsigned magnetic field contained in active regions, for the Zeeman broadening, they produce in NIR atomic lines of various Land\'e factors. This analysis gave a magnetic field of 1.9 kG distributed over 15\% of the stellar surface, corresponding to B$f$ value of 290$\pm$58 G. These values are similar to other magnetic field measurements of stars in the same domain of rotation period and mass \citep{cranmer2011,see2019,shulyak2019,flores2019}. 

In addition, Stokes V profiles for \hd189\ are detected and are compatible with what was found earlier on this star in optical data, with typical values of the longitudinal field of a few Gauss and a median error bar of 1.7 Gauss. There is a mild correlation (Figure \ref{BlBf}) between the longitudinal field obtained with the polarized spectrum, and sensitive to the field large-scale geometry, and the small-scale field flux as measured in the intensity profile of FeI lines through Zeeman broadening. Both properties of the magnetic field of \hd189 thus show significant variations in this data set. The relationship between the small-scale and large-scale magnetic characteristics could be further investigated and modeled, but this preliminary analysis shows that very high S/N spectra ($> 250$ per pixel in the $H$ band) are necessary for this, especially for weak fields. 

While spectropolarimetry of M stars will be greatly improved by the use of \spirou\ compared to optical spectropolarimeters, both domains are seemingly equivalent for early-K stars observed at similar exposure times, and probably reversed for hotter stars, where the line list and relative flux are both privileged in the optical. 

Surface activity of the parent star can, in addition, be examined in the Pa$\beta$ and HeI triplet lines. In the specific data set analyzed here, no significant rotationally modulated variation of these lines could be observed. It would be interesting in a future campaign to relate NIR variability indices to optical chromospheric lines such as H$\alpha$ and the CaII UV doublet or red triplet. This comparison was made for M stars with CARMENES and showed that Pa$\beta$ and HeI variability was not related to optical chromospheric variability in their sample \citep{schoefer2019}. Our data set did not reveal significant activity in these lines, which may be an indication that it is similar for K stars and for M dwarfs.

It would also be interesting to obtain quasi-simultaneous polarimetric information in the optical (e.g., with TBL/NARVAL) and NIR with \spirou. The wealth of information that spectropolarimetry and high-precision velocimetry simultaneously offer for these types of systems indeed allows a complete survey of star and planet properties in the time domain. This is key to understanding evolutions and interactions, especially in close-in planetary systems.

\begin{acknowledgements}{ The authors wish to recognize and acknowledge the very significant cultural role and reverence that the summit of Maunakea has always had within the indigenous Hawaiian community. We are most fortunate to have the opportunity to conduct observations from this mountain. JFD acknowledges funding by the European Research Council (ERC) under the H2020 research \& innovation programme (grant agreement \#740651 NewWorlds). We acknowledge funding from ANR of France under contract number ANR-18-CE31-0019 (SPlaSH). X.D. and G.G. acknowledge support from the French National Research Agency in the framework of the Investissements d'Avenir program (ANR-15-IDEX-02), through the funding of the ”Origin of Life” project of the Univ. Grenoble-Alpes. This work has been carried out within the framework of the NCCR PlanetS supported by the Swiss National Science Foundation. The authors are very grateful to internal referees, Silvia Alencar and Pedro Figueira, for their great help in improving the manuscript. This work has been done in the framework of the \spirou\ Legacy Survey. We also wish to thank the anonymous referee for his/her good suggestions that helped improve the manuscript.
}
\end{acknowledgements}

\bibliographystyle{aa}
\bibliography{references}

\begin{thebibliography}{98}
\expandafter\ifx\csname natexlab\endcsname\relax\def\natexlab#1{#1}\fi

\bibitem[{{Alonso-Floriano} {et~al.}(2019){Alonso-Floriano}, {Snellen},
  {Czesla}, {Bauer}, {Salz}, {Lamp{\'o}n}, {Lara}, {Nagel},
  {L{\'o}pez-Puertas}, {Nortmann}, {S{\'a}nchez-L{\'o}pez}, {Sanz-Forcada},
  {Caballero}, {Reiners}, {Ribas}, {Quirrenbach}, {Amado}, {Aceituno},
  {Anglada-Escud{\'e}}, {B{\'e}jar}, {Brinkm{\"o}ller}, {Hatzes}, {Henning},
  {Kaminski}, {K{\"u}rster}, {Labarga}, {Montes}, {Pall{\'e}}, {Schmitt}, \&
  {Zapatero Osorio}}]{alonsofloriano2019}
{Alonso-Floriano}, F.~J., {Snellen}, I.~A.~G., {Czesla}, S., {et~al.} 2019,
  \aap, 629, A110

\bibitem[{{Artigau} {et~al.}(2014){Artigau}, {Astudillo-Defru}, {Delfosse},
  {Bouchy}, {Bonfils}, {Lovis}, {Pepe}, {Moutou}, {Donati}, {Doyon}, \&
  {Malo}}]{artigau14}
{Artigau}, {\'E}., {Astudillo-Defru}, N., {Delfosse}, X., {et~al.} 2014, in
  Society of Photo-Optical Instrumentation Engineers (SPIE) Conference Series,
  Vol. 9149, \procspie, 914905

\bibitem[{{Artigau} {et~al.}(2018){Artigau}, {Saint-Antoine}, {L{\'e}vesque},
  {Vall{\'e}e}, {Doyon}, {Hernandez}, \& {Moutou}}]{artigau2018}
{Artigau}, {\'E}., {Saint-Antoine}, J., {L{\'e}vesque}, P.-L., {et~al.} 2018,
  in Society of Photo-Optical Instrumentation Engineers (SPIE) Conference
  Series, Vol. 10709, \procspie, 107091P

\bibitem[{{Astudillo-Defru} {et~al.}(2015){Astudillo-Defru}, {Bonfils},
  {Delfosse}, {S{\'e}gransan}, {Forveille}, {Bouchy}, {Gillon}, {Lovis},
  {Mayor}, {Neves}, {Pepe}, {Perrier}, {Queloz}, {Rojo}, {Santos}, \&
  {Udry}}]{astudillo2015}
{Astudillo-Defru}, N., {Bonfils}, X., {Delfosse}, X., {et~al.} 2015, \aap, 575,
  A119

\bibitem[{{Bagnulo} {et~al.}(2009){Bagnulo}, {Landolfi}, {Landstreet}, {Land i
  Degl'Innocenti}, {Fossati}, \& {Sterzik}}]{bagnulo2009}
{Bagnulo}, S., {Landolfi}, M., {Landstreet}, J.~D., {et~al.} 2009, \pasp, 121,
  993

\bibitem[{{Bakos} {et~al.}(2006{\natexlab{a}}){Bakos}, {Knutson}, {Pont},
  {Moutou}, {Charbonneau}, {Shporer}, {Bouchy}, {Everett}, {Hergenrother},
  {Latham}, {Mayor}, {Mazeh}, {Noyes}, {Queloz}, {P{\'a}l}, \&
  {Udry}}]{bakos2006b}
{Bakos}, G.~{\'A}., {Knutson}, H., {Pont}, F., {et~al.} 2006{\natexlab{a}},
  \apj, 650, 1160

\bibitem[{{Bakos} {et~al.}(2006{\natexlab{b}}){Bakos}, {P{\'a}l}, {Latham},
  {Noyes}, \& {Stefanik}}]{bakos2006}
{Bakos}, G.~{\'A}., {P{\'a}l}, A., {Latham}, D.~W., {Noyes}, R.~W., \&
  {Stefanik}, R.~P. 2006{\natexlab{b}}, \apjl, 641, L57

\bibitem[{{Baluev} {et~al.}(2019){Baluev}, {Sokov}, {Jones}, {Shaidulin},
  {Sokova}, {Nielsen}, {Benni}, {Schneiter}, {Villarreal D'Angelo},
  {Fern{\'a}ndez-Laj{\'u}s}, {Di Sisto}, {Ba{\textcommabelow s}t{\"u}rk},
  {Bretton}, {Wunsche}, {Hentunen}, {Shadick}, {Jongen}, {Kang}, {Kim},
  {Pak{\v{s}}tien{\"A}--}, {Qvam}, {Knight}, {Guerra}, {Marchini}, {Salvaggio},
  {Papini}, {Evans}, {Salisbury}, {Garcia}, {Molina}, {Garlitz}, {Esseiva},
  {Ogmen}, {Karavaev}, {Rusov}, {Ibrahimov}, \& {Karimov}}]{baluev2019}
{Baluev}, R.~V., {Sokov}, E.~N., {Jones}, H.~R.~A., {et~al.} 2019, \mnras, 490,
  1294

\bibitem[{{Baranne} {et~al.}(1996){Baranne}, {Queloz}, {Mayor}, {Adrianzyk},
  {Knispel}, {Kohler}, {Lacroix}, {Meunier}, {Rimbaud}, \& {Vin}}]{baranne}
{Baranne}, A., {Queloz}, D., {Mayor}, M., {et~al.} 1996, \aaps, 119, 373

\bibitem[{{Barrick} {et~al.}(2018){Barrick}, {Donati}, {Baratchart}, {Moutou},
  {Vermeulen}, {Ho}, {Larrieu}, {Par{\`e}s}, {Dupieux}, {Wang}, \&
  {Yan}}]{barrick2018}
{Barrick}, G., {Donati}, J.-F., {Baratchart}, S., {et~al.} 2018, in Society of
  Photo-Optical Instrumentation Engineers (SPIE) Conference Series, Vol. 10702,
  \procspie, 1070268

\bibitem[{{Bauer} {et~al.}(2015){Bauer}, {Zechmeister}, \&
  {Reiners}}]{bauer2015}
{Bauer}, F.~F., {Zechmeister}, M., \& {Reiners}, A. 2015, \aap, 581, A117

\bibitem[{{Becerril} {et~al.}(2017){Becerril}, {Mirabet}, {Lizon}, {Calvo},
  {Abril}, {C{\'a}rdenas}, {Ferro}, {Morales}, {P{\'e}rez}, {Ram{\'o}n},
  {S{\'a}nchez-Carrasco}, {Quirrenbach}, {Amado}, {Ribas}, {Reiners},
  {Caballero}, {Seifert}, \& {Herranz}}]{Beceril2017}
{Becerril}, S., {Mirabet}, E., {Lizon}, J.~L., {et~al.} 2017, in Materials
  Science and Engineering Conference Series, Vol. 278, Materials Science and
  Engineering Conference Series, 012191

\bibitem[{{Benatti} {et~al.}(2017){Benatti}, {Desidera}, {Damasso},
  {Malavolta}, {Lanza}, {Biazzo}, {Bonomo}, {Claudi}, {Marzari}, {Poretti},
  {Gratton}, {Micela}, {Pagano}, {Piotto}, {Sozzetti}, {Boccato}, {Cosentino},
  {Covino}, {Maggio}, {Molinari}, {Smareglia}, {Affer}, {Andreuzzi},
  {Bignamini}, {Borsa}, {di Fabrizio}, {Esposito}, {Martinez Fiorenzano},
  {Messina}, {Giacobbe}, {Harutyunyan}, {Knapic}, {Maldonado}, {Masiero},
  {Nascimbeni}, {Pedani}, {Rainer}, {Scandariato}, \& {Silvotti}}]{benatti}
{Benatti}, S., {Desidera}, S., {Damasso}, M., {et~al.} 2017, \aap, 599, A90

\bibitem[{{Birkby} {et~al.}(2013){Birkby}, {de Kok}, {Brogi}, {de Mooij},
  {Schwarz}, {Albrecht}, \& {Snellen}}]{birkby2013}
{Birkby}, J.~L., {de Kok}, R.~J., {Brogi}, M., {et~al.} 2013, \mnras, 436, L35

\bibitem[{{Boisse} {et~al.}(2009){Boisse}, {Moutou}, {Vidal-Madjar}, {Bouchy},
  {Pont}, {H{\'e}brard}, {Bonfils}, {Croll}, {Delfosse}, {Desort}, {Forveille},
  {Lagrange}, {Loeillet}, {Lovis}, {Matthews}, {Mayor}, {Pepe}, {Perrier},
  {Queloz}, {Rowe}, {Santos}, {S{\'e}gransan}, \& {Udry}}]{boisse2009}
{Boisse}, I., {Moutou}, C., {Vidal-Madjar}, A., {et~al.} 2009, \aap, 495, 959

\bibitem[{{Bouchy} {et~al.}(2001){Bouchy}, {Pepe}, \& {Queloz}}]{bouchy2001}
{Bouchy}, F., {Pepe}, F., \& {Queloz}, D. 2001, \aap, 374, 733

\bibitem[{{Bouchy} {et~al.}(2005){Bouchy}, {Udry}, {Mayor}, {Moutou}, {Pont},
  {Iribarne}, {da Silva}, {Ilovaisky}, {Queloz}, {Santos}, {S{\'e}gransan}, \&
  {Zucker}}]{bouchy2005}
{Bouchy}, F., {Udry}, S., {Mayor}, M., {et~al.} 2005, \aap, 444, L15

\bibitem[{{Bourrier} {et~al.}(2020){Bourrier}, {Wheatley}, {Lecavelier des
  Etangs}, {King}, {Louden}, {Ehrenreich}, {Fares}, {Helling}, {Llama},
  {Jardine}, \& {Vidotto}}]{bourrier2020}
{Bourrier}, V., {Wheatley}, P.~J., {Lecavelier des Etangs}, A., {et~al.} 2020,
  \mnras, 240

\bibitem[{{Boyajian} {et~al.}(2015){Boyajian}, {von Braun}, {Feiden}, {Huber},
  {Basu}, {Demarque}, {Fischer}, {Schaefer}, {Mann}, {White}, {Maestro},
  {Brewer}, {Lamell}, {Spada}, {L{\'o}pez-Morales}, {Ireland}, {Farrington},
  {van Belle}, {Kane}, {Jones}, {ten Brummelaar}, {Ciardi}, {McAlister},
  {Ridgway}, {Goldfinger}, {Turner}, \& {Sturmann}}]{boyajian2015}
{Boyajian}, T., {von Braun}, K., {Feiden}, G.~A., {et~al.} 2015, \mnras, 447,
  846

\bibitem[{{Brogi} {et~al.}(2016){Brogi}, {de Kok}, {Albrecht}, {Snellen},
  {Birkby}, \& {Schwarz}}]{brogi2016}
{Brogi}, M., {de Kok}, R.~J., {Albrecht}, S., {et~al.} 2016, \apj, 817, 106

\bibitem[{{Brogi} {et~al.}(2018){Brogi}, {Giacobbe}, {Guilluy}, {de Kok},
  {Sozzetti}, {Mancini}, \& {Bonomo}}]{brogi2018}
{Brogi}, M., {Giacobbe}, P., {Guilluy}, G., {et~al.} 2018, \aap, 615, A16

\bibitem[{{Brogi} \& {Line}(2019)}]{brogi2019}
{Brogi}, M. \& {Line}, M.~R. 2019, \aj, 157, 114

\bibitem[{{Castelli} \& {Kurucz}(2003)}]{castelli2003}
{Castelli}, F. \& {Kurucz}, R.~L. 2003, in IAU Symposium, Vol. 210, Modelling
  of Stellar Atmospheres, ed. N.~{Piskunov}, W.~W. {Weiss}, \& D.~F. {Gray},
  A20

\bibitem[{{Cegla} {et~al.}(2016){Cegla}, {Lovis}, {Bourrier}, {Beeck},
  {Watson}, \& {Pepe}}]{cegla2016}
{Cegla}, H.~M., {Lovis}, C., {Bourrier}, V., {et~al.} 2016, \aap, 588, A127

\bibitem[{{Challita} {et~al.}(2018){Challita}, {Reshetov}, {Baratchart},
  {Barrick}, {Carmona}, {Donati}, {Micheau}, {Moutou}, {Vall{\'e}e}, {Belot},
  {Gallou}, {Par{\`e}s}, {Rabou}, {Thibault}, {Kouach}, {Lacombe},
  {Saddlemyer}, \& {Striebig}}]{challita18}
{Challita}, Z., {Reshetov}, V., {Baratchart}, S., {et~al.} 2018, in Society of
  Photo-Optical Instrumentation Engineers (SPIE) Conference Series, Vol. 10702,
  \procspie, 1070262

\bibitem[{{Claudi} {et~al.}(2017){Claudi}, {Benatti}, {Carleo}, {Ghedina},
  {Guerra}, {Micela}, {Molinari}, {Oliva}, {Rainer}, {Tozzi}, {Baffa},
  {Baruffolo}, {Buchschacher}, {Cecconi}, {Cosentino}, {Fantinel}, {Fini},
  {Ghinassi}, {Giani}, {Gonzalez}, {Gonzalez}, {Gratton}, {Harutyunyan},
  {Hernandez}, {Lodi}, {Malavolta}, {Maldonado}, {Origlia}, {Sanna}, {Sanjuan},
  {Scuderi}, {Seemann}, {Sozzetti}, {Perez Ventura}, {Hernandez Diaz}, {Galli},
  {Gonzalez}, {Riverol}, \& {Riverol}}]{Claudi2017}
{Claudi}, R., {Benatti}, S., {Carleo}, I., {et~al.} 2017, European Physical
  Journal Plus, 132, 364

\bibitem[{{Cloutier} {et~al.}(2018){Cloutier}, {Doyon}, {Bouchy}, \&
  {H{\'e}brard}}]{cloutier2018}
{Cloutier}, R., {Doyon}, R., {Bouchy}, F., \& {H{\'e}brard}, G. 2018, \aj, 156,
  82

\bibitem[{{Collier Cameron} {et~al.}(2010){Collier Cameron}, {Bruce}, {Miller},
  {Triaud}, \& {Queloz}}]{cameron2010}
{Collier Cameron}, A., {Bruce}, V.~A., {Miller}, G.~R.~M., {Triaud},
  A.~H.~M.~J., \& {Queloz}, D. 2010, \mnras, 403, 151

\bibitem[{{Connes}(1985)}]{connes1985}
{Connes}, P. 1985, \apss, 110, 211

\bibitem[{{Cranmer} \& {Saar}(2011)}]{cranmer2011}
{Cranmer}, S.~R. \& {Saar}, S.~H. 2011, \apj, 741, 54

\bibitem[{{Croll} {et~al.}(2007){Croll}, {Matthews}, {Rowe}, {Gladman},
  {Miller-Ricci}, {Sasselov}, {Walker}, {Kuschnig}, {Lin}, {Guenther},
  {Moffat}, {Rucinski}, \& {Weiss}}]{croll2007}
{Croll}, B., {Matthews}, J.~M., {Rowe}, J.~F., {et~al.} 2007, \apj, 671, 2129

\bibitem[{{Cuillandre} {et~al.}(2016){Cuillandre}, {Magnier}, {Sabin}, \&
  {Mahoney}}]{cuillandre}
{Cuillandre}, J.~C., {Magnier}, E., {Sabin}, D., \& {Mahoney}, B. 2016,
  Astronomical Society of the Pacific Conference Series, Vol. 503, {SkyProbe:
  Real-Time Precision Monitoring in the Optical of the Absolute Atmospheric
  Absorption on the Telescope Science and Calibration Fields}, ed.
  S.~{Deustua}, S.~{Allam}, D.~{Tucker}, \& J.~A. {Smith}, 233

\bibitem[{{Dalal} {et~al.}(2019){Dalal}, {H{\'e}brard}, {Lecavelier des
  {\'E}tangs}, {Petit}, {Bourrier}, {Laskar}, {K{\"o}nig}, \&
  {Correia}}]{dalal2019}
{Dalal}, S., {H{\'e}brard}, G., {Lecavelier des {\'E}tangs}, A., {et~al.} 2019,
  \aap, 631, A28

\bibitem[{{Donati} {et~al.}(2018){Donati}, {Kouach}, {Lacombe}, {Baratchart},
  {Doyon}, {Delfosse}, {Artigau}, {Moutou}, {H{\'e}brard}, {Bouchy}, {Bouvier},
  {Alencar}, {Saddlemyer}, {Par{\`e}s}, {Rabou}, {Micheau}, {Dolon}, {Barrick},
  {Hernandez}, {Wang}, {Reshetov}, {Striebig}, {Challita}, {Carmona},
  {Tibault}, {Martioli}, {Figueira}, {Boisse}, \& {Pepe}}]{donati2018}
{Donati}, J.-F., {Kouach}, D., {Lacombe}, M., {et~al.} 2018, {SPIRou: A NIR
  Spectropolarimeter/High-Precision Velocimeter for the CFHT}, 107

\bibitem[{{Donati} {et~al.}(2020){Donati}, {Kouach}, {Moutou}, {Doyon},
  {Delfosse}, \& {Artigau}}]{donati2020}
{Donati}, J.-F., {Kouach}, D., {Moutou}, C., {et~al.} 2020, \mnras, in press

\bibitem[{{Donati} {et~al.}(1997){Donati}, {Semel}, {Carter}, {Rees}, \&
  {Collier Cameron}}]{donati97}
{Donati}, J.~F., {Semel}, M., {Carter}, B.~D., {Rees}, D.~E., \& {Collier
  Cameron}, A. 1997, \mnras, 291, 658

\bibitem[{{Dorn} {et~al.}(2016){Dorn}, {Follert}, {Bristow}, {Cumani},
  {Eschbaumer}, {Grunhut}, {Haimerl}, {Hatzes}, {Heiter}, {Hinterschuster},
  {Ives}, {Jung}, {Kerber}, {Klein}, {Lavail}, {Lizon}, {L{\"o}winger},
  {Molina-Conde}, {Nicholson}, {Marquart}, {Oliva}, {Origlia}, {Pasquini},
  {Paufique}, {Piskunov}, {Reiners}, {Seemann}, {Stegmeier}, {Stempels}, \&
  {Tordo}}]{Dorn2016}
{Dorn}, R.~J., {Follert}, R., {Bristow}, P., {et~al.} 2016, Society of
  Photo-Optical Instrumentation Engineers (SPIE) Conference Series, Vol. 9908,
  {The ``+'' for CRIRES: enabling better science at infrared wavelength and
  high spectral resolution at the ESO VLT}, 99080I

\bibitem[{{Doyle} {et~al.}(2013){Doyle}, {Smalley}, {Maxted}, {Anderson},
  {Cameron}, {Gillon}, {Hellier}, {Pollacco}, {Queloz}, {Triaud}, \&
  {West}}]{Doyle2013}
{Doyle}, A.~P., {Smalley}, B., {Maxted}, P.~F.~L., {et~al.} 2013, \mnras, 428,
  3164

\bibitem[{{Dumusque}(2014)}]{dumusque2014b}
{Dumusque}, X. 2014, \apj, 796, 133

\bibitem[{{Dumusque} {et~al.}(2014){Dumusque}, {Boisse}, \&
  {Santos}}]{dumusque2014a}
{Dumusque}, X., {Boisse}, I., \& {Santos}, N.~C. 2014, \apj, 796, 132

\bibitem[{{Fares} {et~al.}(2017){Fares}, {Bourrier}, {Vidotto}, {Moutou},
  {Jardine}, {Zarka}, {Helling}, {Lecavelier des Etangs}, {Llama}, {Louden},
  {Wheatley}, \& {Ehrenreich}}]{fares17}
{Fares}, R., {Bourrier}, V., {Vidotto}, A.~A., {et~al.} 2017, \mnras, 471, 1246

\bibitem[{{Fares} {et~al.}(2010){Fares}, {Donati}, {Moutou}, {Jardine},
  {Grie{\ss}meier}, {Zarka}, {Shkolnik}, {Bohlender}, {Catala}, \& {Collier
  Cameron}}]{fares10}
{Fares}, R., {Donati}, J.~F., {Moutou}, C., {et~al.} 2010, \mnras, 406, 409

\bibitem[{{Figueira} {et~al.}(2016){Figueira}, {Adibekyan}, {Oshagh}, {Neal},
  {Rojas-Ayala}, {Lovis}, {Melo}, {Pepe}, {Santos}, \&
  {Tsantaki}}]{figueira2016}
{Figueira}, P., {Adibekyan}, V.~Z., {Oshagh}, M., {et~al.} 2016, \aap, 586,
  A101

\bibitem[{{Flores} {et~al.}(2019){Flores}, {Connelley}, {Reipurth}, \&
  {Boogert}}]{flores2019}
{Flores}, C., {Connelley}, M.~S., {Reipurth}, B., \& {Boogert}, A. 2019, \apj,
  882, 75

\bibitem[{{Folsom} {et~al.}(2016){Folsom}, {Petit}, {Bouvier}, {L{\`e}bre},
  {Amard}, {Palacios}, {Morin}, {Donati}, {Jeffers}, {Marsden}, \&
  {Vidotto}}]{Folsom2016}
{Folsom}, C.~P., {Petit}, P., {Bouvier}, J., {et~al.} 2016, \mnras, 457, 580

\bibitem[{{Foreman-Mackey} {et~al.}(2013){Foreman-Mackey}, {Hogg}, {Lang}, \&
  {Goodman}}]{foreman2013}
{Foreman-Mackey}, D., {Hogg}, D.~W., {Lang}, D., \& {Goodman}, J. 2013, \pasp,
  125, 306

\bibitem[{{Galland} {et~al.}(2005){Galland}, {Lagrange}, {Udry}, {Chelli},
  {Pepe}, {Queloz}, {Beuzit}, \& {Mayor}}]{galland2005}
{Galland}, F., {Lagrange}, A.~M., {Udry}, S., {et~al.} 2005, \aap, 443, 337

\bibitem[{{Gustafsson} {et~al.}(2008){Gustafsson}, {Edvardsson}, {Eriksson},
  {J{\o}rgensen}, {Nordlund}, \& {Plez}}]{Gustafsson2008-MARCS-grid}
{Gustafsson}, B., {Edvardsson}, B., {Eriksson}, K., {et~al.} 2008, \aap, 486,
  951

\bibitem[{{Hayek} {et~al.}(2012){Hayek}, {Sing}, {Pont}, \&
  {Asplund}}]{hayek2012}
{Hayek}, W., {Sing}, D., {Pont}, F., \& {Asplund}, M. 2012, \aap, 539, A102

\bibitem[{{H{\'e}brard} {et~al.}(2014){H{\'e}brard}, {Donati}, {Delfosse},
  {Morin}, {Boisse}, {Moutou}, \& {H{\'e}brard}}]{hebrard2014}
{H{\'e}brard}, {\'E}.~M., {Donati}, J.~F., {Delfosse}, X., {et~al.} 2014,
  \mnras, 443, 2599

\bibitem[{{H{\'e}brard} {et~al.}(2016){H{\'e}brard}, {Donati}, {Delfosse},
  {Morin}, {Moutou}, \& {Boisse}}]{hebrard2016}
{H{\'e}brard}, {\'E}.~M., {Donati}, J.~F., {Delfosse}, X., {et~al.} 2016,
  \mnras, 461, 1465

\bibitem[{{Henry} \& {Winn}(2008)}]{henry2008}
{Henry}, G.~W. \& {Winn}, J.~N. 2008, \aj, 135, 68

\bibitem[{{Hobson} {et~al.}(2020){Hobson}, {Bouchy}, {Cook}, {Artigau},
  {Moutou}, \& {Boisse}}]{hobson2020}
{Hobson}, M., {Bouchy}, F., {Cook}, N., {et~al.} 2020, \aap, submitted

\bibitem[{{Horne}(1986)}]{horne86}
{Horne}, K. 1986, \pasp, 98, 609

\bibitem[{{Hussain} {et~al.}(2016){Hussain}, {Alvarado-G{\'o}mez}, {Grunhut},
  {Donati}, {Alecian}, {Oksala}, {Morin}, {Fares}, {Jardine}, {Drake}, {Cohen},
  {Matt}, {Petit}, {Redfield}, \& {Walter}}]{hussain2016}
{Hussain}, G.~A.~J., {Alvarado-G{\'o}mez}, J.~D., {Grunhut}, J., {et~al.} 2016,
  \aap, 585, A77

\bibitem[{{Kotani} {et~al.}(2018){Kotani}, {Tamura}, {Nishikawa}, {Ueda},
  {Kuzuhara}, {Omiya}, {Hashimoto}, {Ishizuka}, {Hirano}, {Suto}, {Kurokawa},
  {Kokubo}, {Mori}, {Tanaka}, {Kashiwagi}, {Konishi}, {Kudo}, {Sato},
  {Jacobson}, {Hodapp}, {Hall}, {Aoki}, {Usuda}, {Nishiyama}, {Nakajima},
  {Ikeda}, {Yamamuro}, {Morino}, {Baba}, {Hosokawa}, {Ishikawa}, {Narita},
  {Kokubo}, {Hayano}, {Izumiura}, {Kambe}, {Kusakabe}, {Kwon}, {Ikoma}, {Hori},
  {Genda}, {Fukui}, {Fujii}, {Kawahara}, {Olivier}, {Jovanovic}, {Harakawa},
  {Hayashi}, {Hidai}, {Machida}, {Matsuo}, {Nagata}, {Ogihara}, {Takami},
  {Takato}, {Terada}, \& {Oh}}]{kotani2018}
{Kotani}, T., {Tamura}, M., {Nishikawa}, J., {et~al.} 2018, in Society of
  Photo-Optical Instrumentation Engineers (SPIE) Conference Series, Vol. 10702,
  \procspie, 1070211

\bibitem[{{Kupka} {et~al.}(1999){Kupka}, {Piskunov}, {Ryabchikova}, {Stempels},
  \& {Weiss}}]{Kupka1999-VALD}
{Kupka}, F., {Piskunov}, N., {Ryabchikova}, T.~A., {Stempels}, H.~C., \&
  {Weiss}, W.~W. 1999, \aaps, 138, 119

\bibitem[{{Kurucz}(2014)}]{Kurucz-K14}
{Kurucz}, R.~L. 2014, Robert L. Kurucz on-line database of observed and
  predicted atomic transitions

\bibitem[{{Landstreet}(1988)}]{Landstreet1988}
{Landstreet}, J.~D. 1988, \apj, 326, 967

\bibitem[{{Lanza} {et~al.}(2011){Lanza}, {Boisse}, {Bouchy}, {Bonomo}, \&
  {Moutou}}]{lanza2011}
{Lanza}, A.~F., {Boisse}, I., {Bouchy}, F., {Bonomo}, A.~S., \& {Moutou}, C.
  2011, \aap, 533, A44

\bibitem[{{Lavail} {et~al.}(2017){Lavail}, {Kochukhov}, {Hussain}, {Alecian},
  {Herczeg}, \& {Johns-Krull}}]{lavail2017}
{Lavail}, A., {Kochukhov}, O., {Hussain}, G.~A.~J., {et~al.} 2017, \aap, 608,
  A77

\bibitem[{{Mahadevan} {et~al.}(2014){Mahadevan}, {Ramsey}, {Terrien},
  {Halverson}, {Roy}, {Hearty}, {Levi}, {Stefansson}, {Robertson}, {Bender},
  {Schwab}, \& {Nelson}}]{mahadevan2014}
{Mahadevan}, S., {Ramsey}, L.~W., {Terrien}, R., {et~al.} 2014, Society of
  Photo-Optical Instrumentation Engineers (SPIE) Conference Series, Vol. 9147,
  {The Habitable-zone Planet Finder: A status update on the development of a
  stabilized fiber-fed near-infrared spectrograph for the for the Hobby-Eberly
  telescope}, 91471G

\bibitem[{{Marcy} \& {Butler}(1992)}]{marcy}
{Marcy}, G.~W. \& {Butler}, R.~P. 1992, \pasp, 104, 270

\bibitem[{{Mayo} {et~al.}(2019){Mayo}, {Rajpaul}, {Buchhave}, {Dressing},
  {Mortier}, {Zeng}, {Fortenbach}, {Aigrain}, {Bonomo}, {Collier Cameron},
  {Charbonneau}, {Coffinet}, {Cosentino}, {Damasso}, {Dumusque}, {Martinez
  Fiorenzano}, {Haywood}, {Latham}, {L{\'o}pez-Morales}, {Malavolta}, {Micela},
  {Molinari}, {Pearce}, {Pepe}, {Phillips}, {Piotto}, {Poretti}, {Rice},
  {Sozzetti}, \& {Udry}}]{mayo2019}
{Mayo}, A.~W., {Rajpaul}, V.~M., {Buchhave}, L.~A., {et~al.} 2019, \aj, 158,
  165

\bibitem[{{Mayor} \& {Queloz}(1995)}]{mayor}
{Mayor}, M. \& {Queloz}, D. 1995, \nat, 378, 355

\bibitem[{{Micheau} {et~al.}(2018){Micheau}, {Kouach}, {Donati}, {Gallou},
  {Belot}, {Striebig}, {Baratchart}, {Challita}, {Par{\`e}s}, {Dubois},
  {Lacombe}, {Barrick}, {Bouchy}, \& {Pepe}}]{micheau2018}
{Micheau}, Y., {Kouach}, D., {Donati}, J.-F., {et~al.} 2018, in Society of
  Photo-Optical Instrumentation Engineers (SPIE) Conference Series, Vol. 10702,
  \procspie, 107025R

\bibitem[{{Morin} {et~al.}(2013){Morin}, {Jardine}, {Reiners}, {Shulyak},
  {Beeck}, {Hallinan}, {Hebb}, {Hussain}, {Jeffers}, {Kochukhov}, {Vidotto}, \&
  {Walkowicz}}]{morin2013}
{Morin}, J., {Jardine}, M., {Reiners}, A., {et~al.} 2013, Astronomische
  Nachrichten, 334, 48

\bibitem[{{Moutou} {et~al.}(2007){Moutou}, {Donati}, {Savalle}, {Hussain},
  {Alecian}, {Bouchy}, {Catala}, {Collier Cameron}, {Udry}, \&
  {Vidal-Madjar}}]{moutou07}
{Moutou}, C., {Donati}, J.~F., {Savalle}, R., {et~al.} 2007, \aap, 473, 651

\bibitem[{{Moutou} {et~al.}(2017){Moutou}, {H{\'e}brard}, {Morin}, {Malo},
  {Fouqu{\'e}}, {Torres-Rivas}, {Martioli}, {Delfosse}, {Artigau}, \&
  {Doyon}}]{moutou2017}
{Moutou}, C., {H{\'e}brard}, E.~M., {Morin}, J., {et~al.} 2017, \mnras, 472,
  4563

\bibitem[{{Ohta} {et~al.}(2005){Ohta}, {Taruya}, \& {Suto}}]{ohta2005}
{Ohta}, Y., {Taruya}, A., \& {Suto}, Y. 2005, \apj, 622, 1118

\bibitem[{{Par{\`e}s} {et~al.}(2012){Par{\`e}s}, {Donati}, {Dupieux}, {Gharsa},
  {Micheau}, {Bouye}, {Dubois}, {Gallou}, {Kouach}, {Barrick}, \&
  {Wang}}]{pares12}
{Par{\`e}s}, L., {Donati}, J.~F., {Dupieux}, M., {et~al.} 2012, in Society of
  Photo-Optical Instrumentation Engineers (SPIE) Conference Series, Vol. 8446,
  \procspie, 84462E

\bibitem[{{Pepe} {et~al.}(2002){Pepe}, {Mayor}, {Galland}, {Naef}, {Queloz},
  {Santos}, {Udry}, \& {Burnet}}]{pepe2002}
{Pepe}, F., {Mayor}, M., {Galland}, F., {et~al.} 2002, \aap, 388, 632

\bibitem[{{Pepe} {et~al.}(2014){Pepe}, {Molaro}, {Cristiani}, {Rebolo},
  {Santos}, {Dekker}, {M{\'e}gevand}, {Zerbi}, {Cabral}, {Di Marcantonio},
  {Abreu}, {Affolter}, {Aliverti}, {Allende Prieto}, {Amate}, {Avila},
  {Baldini}, {Bristow}, {Broeg}, {Cirami}, {Coelho}, {Conconi}, {Coretti},
  {Cupani}, {D'Odorico}, {De Caprio}, {Delabre}, {Dorn}, {Figueira}, {Fragoso},
  {Galeotta}, {Genolet}, {Gomes}, {Gonz{\'a}lez Hern{\'a}ndez}, {Hughes},
  {Iwert}, {Kerber}, {Landoni}, {Lizon}, {Lovis}, {Maire}, {Mannetta},
  {Martins}, {Monteiro}, {Oliveira}, {Poretti}, {Rasilla}, {Riva}, {Santana
  Tschudi}, {Santos}, {Sosnowska}, {Sousa}, {Span{\'o}}, {Tenegi}, {Toso},
  {Vanzella}, {Viel}, \& {Zapatero Osorio}}]{pepe2014}
{Pepe}, F., {Molaro}, P., {Cristiani}, S., {et~al.} 2014, Astronomische
  Nachrichten, 335, 8

\bibitem[{{Perruchot} {et~al.}(2018){Perruchot}, {Hobson}, {Bouchy}, {Dolon},
  {Boisse}, {Moreau}, {Sottile}, \& {Wildi}}]{perruchot}
{Perruchot}, S., {Hobson}, M., {Bouchy}, F., {et~al.} 2018, in Society of
  Photo-Optical Instrumentation Engineers (SPIE) Conference Series, Vol. 10702,
  \procspie, 1070265

\bibitem[{{Petersburg} {et~al.}(2020){Petersburg}, {Joel Ong}, {Zhao},
  {Blackman}, {Brewer}, {Buchhave}, {Cabot}, {Davis}, {Jurgenson}, {Leet},
  {McCracken}, {Sawyer}, {Sharov}, {Tronsgaard}, {Szymkowiak}, \&
  {Fischer}}]{petersburg2020}
{Petersburg}, R.~R., {Joel Ong}, J.~M., {Zhao}, L.~L., {et~al.} 2020, \aj, 159,
  187

\bibitem[{{Pont} {et~al.}(2007){Pont}, {Gilliland}, {Moutou}, {Charbonneau},
  {Bouchy}, {Brown}, {Mayor}, {Queloz}, {Santos}, \& {Udry}}]{pont2007}
{Pont}, F., {Gilliland}, R.~L., {Moutou}, C., {et~al.} 2007, \aap, 476, 1347

\bibitem[{{Redfield} {et~al.}(2008){Redfield}, {Endl}, {Cochran}, \&
  {Koesterke}}]{redfield2008}
{Redfield}, S., {Endl}, M., {Cochran}, W.~D., \& {Koesterke}, L. 2008, \apjl,
  673, L87

\bibitem[{{Reiners} {et~al.}(2010){Reiners}, {Bean}, {Huber}, {Dreizler},
  {Seifahrt}, \& {Czesla}}]{reiners2010}
{Reiners}, A., {Bean}, J.~L., {Huber}, K.~F., {et~al.} 2010, \apj, 710, 432

\bibitem[{{Reiners} {et~al.}(2013){Reiners}, {Shulyak}, {Anglada-Escud{\'e}},
  {Jeffers}, {Morin}, {Zechmeister}, {Kochukhov}, \& {Piskunov}}]{reiners2013}
{Reiners}, A., {Shulyak}, D., {Anglada-Escud{\'e}}, G., {et~al.} 2013, \aap,
  552, A103

\bibitem[{{Reiners} \& {Zechmeister}(2020)}]{reiners2020}
{Reiners}, A. \& {Zechmeister}, M. 2020, \apjs, 247, 11

\bibitem[{{Reshetov} {et~al.}(2012){Reshetov}, {Herriot}, {Thibault},
  {D{\'e}saulniers}, {Saddlemyer}, \& {Loop}}]{reshetov2012}
{Reshetov}, V., {Herriot}, G., {Thibault}, S., {et~al.} 2012, Society of
  Photo-Optical Instrumentation Engineers (SPIE) Conference Series, Vol. 8446,
  {Cryogenic mechanical design: SPIROU spectrograph}, 84464E

\bibitem[{{Ryabchikova} {et~al.}(2015){Ryabchikova}, {Piskunov}, {Kurucz},
  {Stempels}, {Heiter}, {Pakhomov}, \& {Barklem}}]{Ryabchikova2015-VALD3}
{Ryabchikova}, T., {Piskunov}, N., {Kurucz}, R.~L., {et~al.} 2015, PhysS, 90,
  054005

\bibitem[{{Salz} {et~al.}(2018){Salz}, {Czesla}, {Schneider}, {Nagel},
  {Schmitt}, {Nortmann}, {Alonso-Floriano}, {L{\'o}pez-Puertas}, {Lamp{\'o}n},
  {Bauer}, {Snellen}, {Pall{\'e}}, {Caballero}, {Yan}, {Chen}, {Sanz-Forcada},
  {Amado}, {Quirrenbach}, {Ribas}, {Reiners}, {B{\'e}jar}, {Casasayas-Barris},
  {Cort{\'e}s-Contreras}, {Dreizler}, {Guenther}, {Henning}, {Jeffers},
  {Kaminski}, {K{\"u}rster}, {Lafarga}, {Lara}, {Molaverdikhani}, {Montes},
  {Morales}, {S{\'a}nchez-L{\'o}pez}, {Seifert}, {Zapatero Osorio}, \&
  {Zechmeister}}]{salz2018}
{Salz}, M., {Czesla}, S., {Schneider}, P.~C., {et~al.} 2018, \aap, 620, A97

\bibitem[{{Sch{\"o}fer} {et~al.}(2019){Sch{\"o}fer}, {Jeffers}, {Reiners},
  {Shulyak}, {Fuhrmeister}, {Johnson}, {Zechmeister}, {Ribas}, {Quirrenbach},
  {Amado}, {Caballero}, {Anglada-Escud{\'e}}, {Bauer}, {B{\'e}jar},
  {Cort{\'e}s-Contreras}, {Dreizler}, {Guenther}, {Kaminski}, {K{\"u}rster},
  {Lafarga}, {Montes}, {Morales}, {Pedraz}, \& {Tal-Or}}]{schoefer2019}
{Sch{\"o}fer}, P., {Jeffers}, S.~V., {Reiners}, A., {et~al.} 2019, \aap, 623,
  A44

\bibitem[{{See} {et~al.}(2019){See}, {Matt}, {Folsom}, {Boro Saikia}, {Donati},
  {Fares}, {Finley}, {H{\'e}brard}, {Jardine}, {Jeffers}, {Lehmann}, {Marsden},
  {Mengel}, {Morin}, {Petit}, {Vidotto}, {Waite}, \& {BCool
  Collaboration}}]{see2019}
{See}, V., {Matt}, S.~P., {Folsom}, C.~P., {et~al.} 2019, \apj, 876, 118

\bibitem[{{Shulyak} {et~al.}(2019){Shulyak}, {Reiners}, {Nagel}, {Tal-Or},
  {Caballero}, {Zechmeister}, {B{\'e}jar}, {Cort{\'e}s-Contreras}, {Martin},
  {Kaminski}, {Ribas}, {Quirrenbach}, {Amado}, {Anglada-Escud{\'e}}, {Bauer},
  {Dreizler}, {Guenther}, {Henning}, {Jeffers}, {K{\"u}rster}, {Lafarga},
  {Montes}, {Morales}, \& {Pedraz}}]{shulyak2019}
{Shulyak}, D., {Reiners}, A., {Nagel}, E., {et~al.} 2019, \aap, 626, A86

\bibitem[{{Sing} {et~al.}(2011){Sing}, {Pont}, {Aigrain}, {Charbonneau},
  {D{\'e}sert}, {Gibson}, {Gilliland}, {Hayek}, {Henry}, {Knutson}, {Lecavelier
  Des Etangs}, {Mazeh}, \& {Shporer}}]{sing2011}
{Sing}, D.~K., {Pont}, F., {Aigrain}, S., {et~al.} 2011, \mnras, 416, 1443

\bibitem[{{Thibault} {et~al.}(2012){Thibault}, {Rabou}, {Donati},
  {Desaulniers}, {Dallaire}, {Artigau}, {Pepe}, {Micheau}, {Vall{\'e}e},
  {Pepe}, {Barrick}, {Reshetov}, {Hernand ez}, {Saddlemyer}, {Pazder},
  {Par{\`e}s}, {Doyon}, {Delfosse}, {Kouach}, \& {Loop}}]{thibaut2012}
{Thibault}, S., {Rabou}, P., {Donati}, J.-F., {et~al.} 2012, Society of
  Photo-Optical Instrumentation Engineers (SPIE) Conference Series, Vol. 8446,
  {SPIRou @ CFHT: spectrograph optical design}, 844630

\bibitem[{{Torres} {et~al.}(2008){Torres}, {Winn}, \& {Holman}}]{torres2008}
{Torres}, G., {Winn}, J.~N., \& {Holman}, M.~J. 2008, \apj, 677, 1324

\bibitem[{{Triaud} {et~al.}(2009){Triaud}, {Queloz}, {Bouchy}, {Moutou},
  {Collier Cameron}, {Claret}, {Barge}, {Benz}, {Deleuil}, {Guillot},
  {H{\'e}brard}, {Lecavelier Des {\'E}tangs}, {Lovis}, {Mayor}, {Pepe}, \&
  {Udry}}]{triaud09}
{Triaud}, A.~H.~M.~J., {Queloz}, D., {Bouchy}, F., {et~al.} 2009, \aap, 506,
  377

\bibitem[{{Valenti} {et~al.}(1995){Valenti}, {Marcy}, \& {Basri}}]{Valenti1995}
{Valenti}, J.~A., {Marcy}, G.~W., \& {Basri}, G. 1995, \apj, 439, 939

\bibitem[{{Vidotto} {et~al.}(2014){Vidotto}, {Gregory}, {Jardine}, {Donati},
  {Petit}, {Morin}, {Folsom}, {Bouvier}, {Cameron}, {Hussain}, {Marsden},
  {Waite}, {Fares}, {Jeffers}, \& {do Nascimento}}]{vidotto2014}
{Vidotto}, A.~A., {Gregory}, S.~G., {Jardine}, M., {et~al.} 2014, \mnras, 441,
  2361

\bibitem[{{Wade} {et~al.}(2001){Wade}, {Bagnulo}, {Kochukhov}, {Landstreet},
  {Piskunov}, \& {Stift}}]{Wade2001}
{Wade}, G.~A., {Bagnulo}, S., {Kochukhov}, O., {et~al.} 2001, \aap, 374, 265

\bibitem[{{Walker} {et~al.}(1995){Walker}, {Walker}, {Irwin}, {Larson}, {Yang},
  \& {Richardson}}]{walker}
{Walker}, G. A.~H., {Walker}, A.~R., {Irwin}, A.~W., {et~al.} 1995, \icarus,
  116, 359

\bibitem[{{Winn} {et~al.}(2007){Winn}, {Holman}, {Henry}, {Roussanova}, {Enya},
  {Yoshii}, {Shporer}, {Mazeh}, {Johnson}, {Narita}, \& {Suto}}]{winn2007}
{Winn}, J.~N., {Holman}, M.~J., {Henry}, G.~W., {et~al.} 2007, \aj, 133, 1828

\bibitem[{{Winn} {et~al.}(2006){Winn}, {Johnson}, {Marcy}, {Butler}, {Vogt},
  {Henry}, {Roussanova}, {Holman}, {Enya}, {Narita}, {Suto}, \&
  {Turner}}]{winn2006}
{Winn}, J.~N., {Johnson}, J.~A., {Marcy}, G.~W., {et~al.} 2006, \apjl, 653, L69

\bibitem[{{Yu} {et~al.}(2019){Yu}, {Donati}, {Grankin}, {Collier Cameron},
  {Moutou}, {Hussain}, {Baruteau}, {Jouve}, \& {MaTYSSE
  Collaboration}}]{yu2019}
{Yu}, L., {Donati}, J.~F., {Grankin}, K., {et~al.} 2019, \mnras, 489, 5556

\bibitem[{{Zechmeister} {et~al.}(2018){Zechmeister}, {Reiners}, {Amado},
  {Azzaro}, {Bauer}, {B{\'e}jar}, {Caballero}, {Guenther}, {Hagen}, {Jeffers},
  {Kaminski}, {K{\"u}rster}, {Launhardt}, {Montes}, {Morales}, {Quirrenbach},
  {Reffert}, {Ribas}, {Seifert}, {Tal-Or}, \& {Wolthoff}}]{zechmeister2018}
{Zechmeister}, M., {Reiners}, A., {Amado}, P.~J., {et~al.} 2018, \aap, 609, A12

\end{thebibliography}

\newpage
\onecolumn
\begin{longtable}{ccc}
\caption{Radial velocity data of \hd189 measured by \spirou\ in the K2-filtered mask.}\\
\label{tab:rvs}
BJD&RV&$\sigma$\\
&km/s&m/s\\
\hline
   2458329.01846& -2.4639&    1.47\\
   2458329.02200& -2.4560&    1.54\\
   2458329.02524& -2.4509&    1.49\\
   2458329.02842& -2.4558&    1.44\\
   2458329.97258& -2.0407&    1.61\\
   2458329.97588& -2.0392&    1.59\\
   2458329.97910& -2.0292&    1.65\\
   2458329.98229& -2.0272&    1.62\\
   2458331.94896& -2.0830&    1.80\\
   2458331.95230& -2.0994&    1.77\\
   2458331.95551& -2.0756&    1.79\\
   2458331.95871& -2.0854&    1.78\\
   2458333.01588& -2.3947&    1.81\\
   2458333.01932& -2.3947&    1.79\\
   2458333.02258& -2.4050&    1.86\\
   2458333.02580& -2.3863&    1.97\\
   2458333.97185& -2.2169&    3.23\\
   2458333.97513& -2.2104&    2.70\\
   2458333.97841& -2.2214&    3.60\\
   2458333.98159& -2.2412&    3.29\\
   2458333.98600& -2.2350&    3.28\\
   2458333.98929& -2.2052&    2.32\\
   2458333.99252& -2.2040&    2.77\\
   2458333.99571& -2.1863&    3.23\\
   2458335.03784& -2.2936&    2.21\\
   2458335.04119& -2.2832&    2.17\\
   2458335.04443& -2.3018&    2.03\\
   2458335.04763& -2.3009&    1.91\\
   2458336.02048& -2.3131&    1.74\\
   2458336.02377& -2.3105&    1.78\\
   2458336.02699& -2.2979&    1.83\\
   2458336.03020& -2.2924&    1.78\\
   2458336.95142& -2.1189&    1.71\\
   2458336.95482& -2.1118&    1.75\\
   2458336.95809& -2.1253&    1.83\\
   2458336.96127& -2.1236&    1.80\\
   2458383.76872& -2.2362&    1.67\\
   2458383.77208& -2.2347&    1.63\\
   2458383.77531& -2.2152&    1.61\\
   2458383.77855& -2.2208&    1.64\\
   2458383.78181& -2.2307&    1.65\\
   2458383.78506& -2.2386&    1.60\\
   2458383.78830& -2.2363&    1.70\\
   2458383.79151& -2.2498&    1.57\\
   2458383.79478& -2.2600&    1.62\\
   2458383.79805& -2.2565&    1.67\\
   2458383.80132& -2.2669&    1.64\\
   2458383.80455& -2.2890&    1.64\\
   2458383.80781& -2.2928&    1.68\\
   2458383.81106& -2.3007&    1.69\\
   2458383.81429& -2.3137&    1.70\\
   2458383.81753& -2.3199&    1.69\\
   2458383.82073& -2.3260&    1.70\\
   2458383.82399& -2.3306&    1.68\\
   2458383.82724& -2.3116&    1.74\\
   2458383.83047& -2.3118&    1.63\\
   2458383.83370& -2.2989&    1.62\\
   2458383.83697& -2.2877&    1.61\\
   2458383.84022& -2.2942&    1.68\\
   2458383.84344& -2.2967&    1.61\\
   2458383.84671& -2.2944&    1.64\\
   2458383.84995& -2.3060&    1.64\\
   2458383.85317& -2.3032&    1.68\\
   2458383.85646& -2.3044&    1.65\\
   2458383.85969& -2.3056&    1.61\\
   2458383.86291& -2.3029&    1.63\\
   2458383.86615& -2.3025&    1.71\\
   2458383.86940& -2.3077&    1.66\\
   2458383.87265& -2.3113&    1.65\\
   2458383.87588& -2.3106&    1.66\\
   2458383.87914& -2.3207&    1.65\\
   2458383.88241& -2.3171&    1.57\\
   2458384.82524& -2.3064&    1.50\\
   2458384.82855& -2.3013&    1.50\\
   2458384.83188& -2.2952&    1.53\\
   2458384.83510& -2.2875&    1.54\\
   2458386.79707& -2.4181&    1.78\\
   2458386.80043& -2.4207&    1.71\\
   2458386.80370& -2.4290&    1.82\\
   2458386.80691& -2.4172&    1.72\\
   2458387.86559& -2.0870&    1.41\\
   2458387.86896& -2.1030&    1.50\\
   2458387.87220& -2.0893&    1.49\\
   2458387.87544& -2.0839&    1.52\\
   2458417.74016& -2.4572&    1.71\\
   2458417.74357& -2.4490&    1.71\\
   2458417.74681& -2.4492&    1.70\\
   2458417.75005& -2.4563&    1.64\\
   2458649.95107& -2.2324&    1.71\\
   2458649.95441& -2.2296&    1.63\\
   2458649.95768& -2.2356&    1.76\\
   2458649.96090& -2.2340&    1.57\\
   2458649.96416& -2.2356&    1.85\\
   2458649.96743& -2.2333&    1.65\\
   2458649.97072& -2.2474&    1.64\\
   2458649.97398& -2.2368&    1.66\\
   2458649.97723& -2.2405&    1.72\\
   2458649.98045& -2.2377&    1.93\\
   2458649.98368& -2.2501&    1.71\\
   2458649.98695& -2.2410&    1.68\\
   2458649.99019& -2.2507&    1.76\\
   2458649.99344& -2.2567&    1.65\\
   2458649.99670& -2.2387&    1.79\\
   2458649.99993& -2.2231&    1.79\\
   2458650.00315& -2.2106&    2.02\\
   2458650.00643& -2.2121&    1.93\\
   2458650.00967& -2.2048&    1.87\\
   2458650.01289& -2.2219&    2.04\\
   2458650.01618& -2.2345&    1.90\\
   2458650.01942& -2.2362&    1.94\\
   2458650.02271& -2.2455&    1.88\\
   2458650.02593& -2.2549&    1.78\\
   2458650.02916& -2.2744&    1.71\\
   2458650.03245& -2.2841&    1.61\\
   2458650.03572& -2.2909&    1.65\\
   2458650.03898& -2.2930&    1.67\\
   2458650.04225& -2.3018&    1.69\\
   2458650.04549& -2.2997&    1.76\\
   2458650.04877& -2.3102&    1.83\\
   2458650.05200& -2.3168&    1.71\\
   2458650.05522& -2.3049&    1.68\\
   2458650.05844& -2.2978&    1.70\\
   2458650.06168& -2.2974&    1.71\\
   2458650.06502& -2.2867&    1.63\\
   2458650.06825& -2.2914&    2.07\\
   2458650.07147& -2.2876&    1.84\\
   2458650.07471& -2.2970&    1.74\\
   2458650.07799& -2.2855&    1.64\\
   2458650.08121& -2.2832&    1.89\\
   2458650.08444& -2.3034&    1.84\\
   2458650.08773& -2.2931&    1.94\\
   2458650.09096& -2.2956&    1.79\\
   2458650.09418& -2.3023&    1.81\\
   2458650.09741& -2.3085&    1.70\\
   2458650.10068& -2.3089&    1.71\\
   2458650.10392& -2.3134&    1.64\\
   2458650.10714& -2.3025&    1.68\\
   2458650.11037& -2.3075&    1.66\\

\hline
\end{longtable}

\end{document}